\documentclass[twocolumn,superscriptaddress,showpacs,aps,amsmath,amssymb,nofootinbib]{revtex4-1}                    %
\usepackage{natbib}
\usepackage{nicefrac}
\usepackage[english]{babel}
\usepackage{graphicx,color}
\usepackage{amsmath,amssymb}
\usepackage{verbatim}
\usepackage{float}
\usepackage{wasysym}
\usepackage{amssymb,graphicx}
\usepackage{epsfig}
\usepackage{psfrag}
\usepackage{dsfont}
\usepackage{amsfonts}
\usepackage{mathrsfs}
\usepackage{multirow}
\usepackage{times}
\usepackage{bm}
\usepackage{hyperref}
\usepackage{xspace}
\usepackage{comment}
\usepackage{scalerel}
\usepackage{soul}
\usepackage[normalem]{ulem}
\hypersetup{
colorlinks=true,        
linkcolor=blue,         
citecolor=cyan,         
}
\usepackage{pifont}
\usepackage{multirow}
\usepackage{verbatim}
\graphicspath{{./}}

\begin{document}
%
\title{Self-consistent treatment of thermal effects in neutron-star post-mergers: \\ observational implications for third-generation gravitational-wave detectors}
\author{Ver\'{o}nica Villa-Ortega}
\affiliation{Instituto Galego de F\'{i}sica de Altas Enerx\'{i}as, Universidade de Santiago de Compostela, 15782 Santiago de Compostela, Galicia, Spain}
 
\author{Ana Lorenzo-Medina}
\affiliation{Instituto Galego de F\'{i}sica de Altas Enerx\'{i}as, Universidade de Santiago de Compostela, 15782 Santiago de Compostela, Galicia, Spain}
\author{Juan Calder\'on~Bustillo}
\affiliation{Instituto Galego de F\'{i}sica de Altas Enerx\'{i}as, Universidade de Santiago de Compostela, 15782 Santiago de Compostela, Galicia, Spain}
\affiliation{Department of Physics, The Chinese University of Hong Kong, Shatin, N.T., Hong Kong}
\author{Milton Ruiz}
 \affiliation{Departament d'Astronomia i Astrofísica, Universitat de València, C/ Dr Moliner 50, 46100, Burjassot (València), Spain} 
\author{Davide Guerra}
\affiliation{Departament d'Astronomia i Astrofísica, Universitat de València, C/ Dr Moliner 50, 46100, Burjassot (València), Spain} 
\author{Pablo Cerdá-Durán}
\affiliation{Departament d'Astronomia i Astrofísica, Universitat de València, C/ Dr Moliner 50, 46100, Burjassot (València), Spain} 
\author{José A. Font}
\affiliation{Departament d'Astronomia i Astrofísica, Universitat de València, C/ Dr Moliner 50, 46100, Burjassot (València), Spain} 
\affiliation{ Observatori Astronòmic, Universitat de València, C/ Catedrático José Beltrán 2, 46980, Paterna (València), Spain}
%
\begin{abstract}
We assess the impact of accurate, self-consistent modelling of thermal effects in neutron-star merger remnants in the context of third-generation 
gravitational-wave detectors. This is done through the usage, in Bayesian model selection experiments, of numerical-relativity simulations of binary 
neutron star (BNS) mergers modelled through: i) nuclear, finite-temperature (or ``tabulated'') equations of state (EoSs), and ii) their  simplifed 
piecewise (or ``hybrid'') representation using seven or ten polytropic pieces. These cover four different EoSs, namely \texttt{SLy4}, \texttt{DD2}, 
\texttt{HShen} and \texttt{LS220}. Our analyses make direct use of the Newman-Penrose scalar $\psi_4$ outputted by numerical simulations. 
Considering a detector network formed by three Cosmic Explorers, we show that differences in the gravitational-wave emission predicted by 
the two models are detectable with a natural logarithmic Bayes Factor $\log{\cal{B}}\geq 5$ at average distances of $d_L \simeq 50$Mpc, 
reaching $d_L \simeq 100$Mpc for source inclinations $\iota \leq 0.8$, regardless of the EoS. This impact is most pronounced for the 
\texttt{HShen} EoS. For low inclinations, only the \texttt{DD2} EoS prevents the detectability of such modelling differences at 
$d_L \simeq 150$Mpc. Our results suggest that the usage a self-consistent treatment of thermal effects is crucial for third-generation gravitational 
wave detectors.
\end{abstract}
\maketitle

%
\section{Introduction}
Mergers of stellar compact binaries, having at least a neutron star as one of the components, lie at the intersection of several very active fields of research in relativistic astrophysics. They are prime targets for multi-messenger observations, being visible in both gravitational wave (GW) and electromagnetic (EM) radiation. The LIGO-Virgo-KAGRA (LVK) Collaboration has reported observations from both types of systems~(see~e.g.~\cite{LIGOScientific:2021djp}). In particular, the detection of the first binary neutron star (BNS) merger, GW170817, along with its post-merger EM emission~\citep{GBM:2017lvd,LIGOScientific:2017zic,LIGOScientific:2017pwl}, has been instrumental to  begin addressing some long-standing questions.  It has placed tight constraints on: i) the equation of state (EoS) at supranuclear densities~\citep{Shibata:2017xdx,Margalit:2017dij,Rezzolla:2017aly,Ruiz:2017due};
ii) the radius and tidal deformability of a spherical neutron star~\citep{Bauswein:2017vtn,Most:2018hfd}; and iii) the speed of GW, which in turn 
disfavours a large class of scalar-tensor theories and other theories predicting varying GW speed~\citep{Ezquiaga:2017ekz,Pardo:2018ipy}. Besides, it also provided an independent measure for the expansion of the Universe~\citep{LIGOScientific:2017adf,Dietrich:2020efo} and the most direct evidence that BNS mergers are progenitors of the central engines of short gamma-ray bursts (sGRBs),
followed by a longer optical transient afterglow known as a
kilonova, powered by the radioactive decay of heavy r-process nuclei~\citep{2010MNRAS.406.2650M,Cowperthwaite:2017,Kasen:2017,Pian:2017}.

Many of those results are drawn from the analysis of the information encoded in the GW signal at the late BNS inspiral stage. While searches have been conducted by the LVK Collaboration~\citep{GW170817-post-merger} no observational evidence of the post-merger GW signal is yet available, as the sensitivity of current detectors at the kHz frequency range is severely hampered by photon shot noise. Estimates on the detectability of the post-merger signal with third-generation detectors have been reported by~\cite{Clark:2016} and~\cite{Miravet-Tenes:2023kte}. 
The BNS inspiral is accurately modelled assuming a cold (zero temperature) EoS.  During this epoch, tidal forces transfer energy and angular momentum from the
orbit to the NS. The energy transferred is primarily converted into gravitational waves  and deformation of the NS structure. However, the internal heating due to these forces is minimal.
On the other hand, during and after merger,
shocks heat up the binary remnant to temperatures $\gtrsim 10~\rm MeV$ adding a thermal pressure support that may change the internal structure of the remnant and its subsequent evolution. It is expected that these thermal effects are encoded in the emitted GWs~\citep{Raithel:2021hye}. Therefore,  observations during the post-merger stage may shed light on the microphysical EoS of NSs at finite temperature.
These observations may become possible with third-generation GW observatories~\citep{Chatziioannou:2017ixj}. 

Advances in our knowledge of the dynamics of BNS mergers and the subsequent post-merger evolution rely on numerical relativity simulations~(for recently reviews see~e.g.~\cite{Baiotti:2016qnr,Shibata:2019wef,Sarin:2020gxb,2020PhRvL.125n1103B,Ruiz:2021gsv}). 
In those simulations NS matter is described using mainly two approaches. The first one is 
a ``hybrid'' approach that assumes that pressure and internal energy can be divided into two contributions~\citep{1993A&A...268..360J,2002A&A...388..917D}: i) a cold part,  computed  through a zero-temperature EoS, $P_{\rm cold}=\kappa_i\,\rho{_0}^{\Gamma_i}$, in a set of intervals in rest-mass density $\rho_0$ (typically referred as a piecewise-polytropic representation of a nuclear EoS~\citep{Read:2008iy}); and ii) a thermal (ideal-gas-like) part to account for shock heating,  $P_{\rm th} =\epsilon_{\rm th}\,(\Gamma_{\rm th}-1)$, which is computed through a thermal  index $\Gamma_{\rm th}$.  Here, $\kappa_i$ and $\Gamma_i$  are the polytropic constant and index, respectively, and $P_{\rm th}$ and $\epsilon_{\rm th}$ the thermal pressure and thermal energy density. Moreover, $\Gamma_{\rm th}$ is a constant ranging between $1$ and $2$~\citep{Constantinou:2015mna}. Above half nuclear saturation density $\Gamma_{\rm th}$ strongly depends on the nucleon effective mass~\citep{Lim:2019ozm}. Hence a constant value of $\Gamma_{\rm th}$ may overestimate the thermal pressure by a few orders of magnitude~\citep{Raithel:2021hye} which, in turn, may induce significant changes in the GW frequencies~\citep{Bauswein:2010dn,Figura:2021bcn}.
The second approach is based on the use of microphysical, finite-temperature EoS tables constructed using ``tabulated'' data from observations and nuclear physics experiments~\citep{Oechslin:2006uk,Bauswein:2010dn,Sekiguchi:2011zd,Fields:2023bhs,Espino:2022mtb,Werneck:2022exo}. This approach provides a self-consistent method
for probing the impact of thermal effects on the fate of the binary remnant, {although it requires the inclusion of other physical phenomena, such as 
neutrino transport~\cite{Zappa:2022rpd}, composition changes in the matter~\cite{Blacker:2023opp}, and induced magnetic viscosity, that  may affect the stability of the binary remnant~\cite{Bamber:2024kfb,Tsokaros:2024wgb}, 
and that, for simplicity, 
we are not taking  into account in our simulations.}
{Note that there are no many or general thermal EoSs 
publicly available~\cite{stellarcollapse}. Some efforts have been made to overcome this limitation~(see~e.g.~\cite{Raithel:2019gws,Mroczek:2024sfp}).  
 Raithel et al.~\cite{Raithel:2019gws} introduced a framework extending cold EoSs to finite temperatures 
and electron fractions using an effective mass and symmetry energy parametrization, achieving $<30\%$ error in 
neutron star merger conditions.  Mroczek et al.~\cite{Mroczek:2024sfp} introduced a controlled expansion method to incorporate thermal effects and arbitrary 
range of charge fractions, from pure neutron to isospin symmetric nuclear matter. This approach reproduces microscopic EoS results up 
to temperatures of $100$ MeV and baryon chemical potentials around $1-2$ times nuclear saturation density with less than $5\%$ error.}

In this work, we assess the importance of a self-consistent (or ``tabulated'') treatment of thermal effects in BNS post-merger 
remnants within a fully Bayesian framework. We use the numerical-relativity waveforms recently obtained by~\cite{Guerra:2025} 
both as ``reference'' signals (or injections) and as recovery templates, considering sources where thermal effects are modeled 
both through a simplified ``hybrid'' approach and a more realistic ``tabulated'' one.  {Our sample of EoSs from the 
StellarCollapse database}~\cite{stellarcollapse} includes the~\texttt{SLy4, DD2, HShen} 
and \texttt{LS220} cases. Given reference data consisting on a synthetic ``tabulated'' signal, we can obtain Bayesian evidences for both the ``tabulated'' and ``hybrid'' waveform models. With this, we evaluate the level to which the difference between the two treatments of thermal effects leads to observable effects in the GW emission. We do this through a novel technique that makes direct use of the Newman-Penrose $\psi_4$ scalar directly outputted by numerical-relativity simulations~\citep{Psi4PE}, as opposed to traditional analyses making use of the GW strain $h(t)$. This prevents the occurrence of systematic numerical errors that might arise when computing the latter from the former through a double time integration~\citep{Pollney_Reissweig}. 
We show that, regardless of the EoS, differences between ``tabulated'' and ``hybrid'' treatments of thermal effects lead to differences in the post-merger GW that are observable by third-generation detectors at source distances
$d_L \leq 50$ Mpc (averaged over the source sky-location). Such distances reach $150$ Mpc for orbital inclinations $\iota < 0.8$, with the exception of BNSs modelled with the \texttt{DD2} EoS. 

On a similar note,  {Fields et al.}~\cite{Fields:2023bhs} also studied the impact of thermal effects on the GWs from BNS mergers by changing the effective masses of neutrons and protons, and thus the specific heat capacity of the binary.  They found that an increased heat capacity results in denser, cooler remnants, which leaves imprints on the GWs. We note that these imprints may be actually due to changes in the nuclei properties triggered by changes in the effective mass. However,
\cite{Fields:2023bhs} claim that the effective mass parameter only affects the kinetic energy of nucleons and has minimal impact on the cold properties of the EoS, and so changes in the GWs should be attributed  {only} to thermal effects.

In addition, \cite{Miravet-Tenes:2023} have recently used the same simulations of~\citet{Guerra:2025} to study the detectability of thermal effects in post-merger signals using \texttt{Bayeswave}~\citep{Cornish:2015}, a Bayesian data-analysis algorithm that reconstructs signals injected into noise through a morphology-independent approach. \cite{Miravet-Tenes:2023} find that differences in the distribution of the main frequency peaks in the spectra in hybrid and tabulated models can be resolved in third-generation detectors up to distances similar to those reported in this work.

\section{Numerical Setup}
\label{sec:Numerical_set}
Our study is based on the recent numerical-relativity simulations performed by~\cite{Guerra:2025}. We briefly describe them here, addressing interested readers to~\cite{Guerra:2025} for full details. 
The binaries consist  of two equal-mass, irrotational NSs modeled by finite-temperature, microphysical EoSs (see Table~\ref{tab:initial_parameters}). The initial temperature is fixed to $T = 0.01\,\rm MeV$.  We use the tables by~\cite{Schneider:2017tfi}, freely available at~\cite{stellarcollapse}. We choose four different EoS that span a suitable range of NS central densities, radii and maximum gravitational masses for irrotational NSs. We also consider a piecewise-polytropic representation of the cold part of these EoSs using a piecewise regression as in~\cite{2021JOSS....6.3859P} with seven (and ten; see Appendix~\ref{Appendix}) polytropic pieces~\citep{Read:2008iy}.
To this cold part we add an ideal-gas part with a constant adiabatic index $\Gamma_{\rm th} = 1.8$ to account for thermal effects.  Initial data are obtained using {\tt LORENE}~\citep{Gourgoulhon:2000nn,tg02}.
%
%
\begin{table}
    \centering
    \begin{center}
        \caption{Summary of the initial properties of the tabulated BNS configurations: 
        We list the EoS, the gravitational mass $M \,[M_\odot]$, the compactness $\mathcal{C}\equiv M/R_{\rm eq}$, the second Love number $k_2$, and the tidal deformability $\Lambda = (2/3)\kappa_2\,\mathcal{C}^{-5}$ for an isolated NS with the same $M$. Here $R_{\rm eq}$ is the equatorial coordinate stellar radius. The last three columns report the ADM mass $M_{\rm ADM}\, [M_\odot]$, the ADM angular momentum $J_{\rm ADM}\, [M_\odot^2]$ and the angular velocity $\Omega \, [\rm krad/s]$.  In all cases the NSs have a rest-mass $M_0=1.4M_\odot$, and an initial temperature of $0.01\rm MeV$. The initial binary coordinate separation is $\sim 44.3\,\rm km$. The hybrid BNS models have similar initial properties (see~\cite{Guerra:2025}). 
        }
    \begin{tabular}{c|ccccccc}
       EoS         & $M$      & $\mathcal{C}$  & $k_2$&  $\Lambda$   &$M_{\rm ADM}$  & $J_{\rm ADM}$ & $\Omega$\\
      \hline
        \texttt{SLy4}      &  1.28   & 0.13    &0.086& 536.09   & 2.54          & 6.63    &  1.77 \\
        \texttt{DD2}        &  1.29   & 0.11   &0.105& 1100.92   & 2.56          & 6.73    &  1.77 \\
        \texttt{HShen}      &  1.30   & 0.10   &0.109& 1805.63 & 2.58          & 6.82    &  1.78 \\
        \texttt{LS220}      &  1.29   & 0.12   &0.106& 915.00   & 2.55          & 6.68    &  1.77 \\
        \end{tabular}
        \label{tab:initial_parameters}
        
    \end{center}
\end{table}
%
\begin{figure*}[t!]
\centering
   \includegraphics[width=0.97\linewidth]{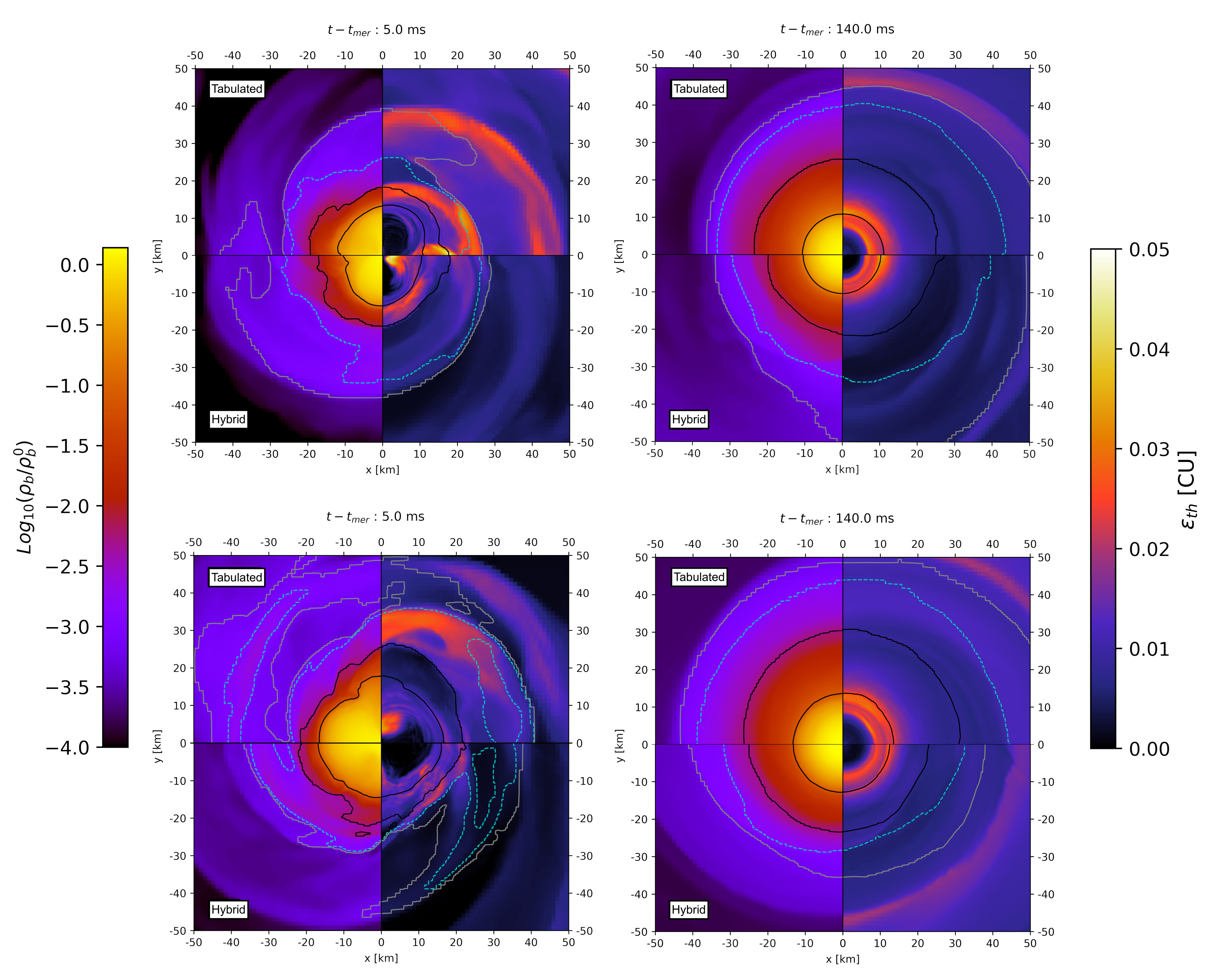}
   \caption{Each panel displays the rest-mass density (left) and thermal specific internal energy 
   (right) of the binary remnant on the equatorial plane at selected times. Those are shown for
   the tabulated (top) and hybrid (bottom) cases of the \texttt{DD2} EoS (top row) and \texttt{HShen} (bottom row) EoS. 
   Isocontours correspond to rest-mass densities~$\rho_0=\{10^{11}, 10^{12},10^{14}\}\,\rm g\,cm^{-3}$. The boundary of the bulk  of the star is displayed as dashed (contour) lines defined
   as regions where $\rho_0=10^{-3}\,{\rho}_{\rm 0, max}$, where ${\rho}_{\rm 0, max}$ is the initial maximum value of the rest-mass density.}
\label{fig:evo_DD2}
\end{figure*}
%
%
%
\begin{figure*}[t!]
\begin{center}
\includegraphics[width=0.95\textwidth]{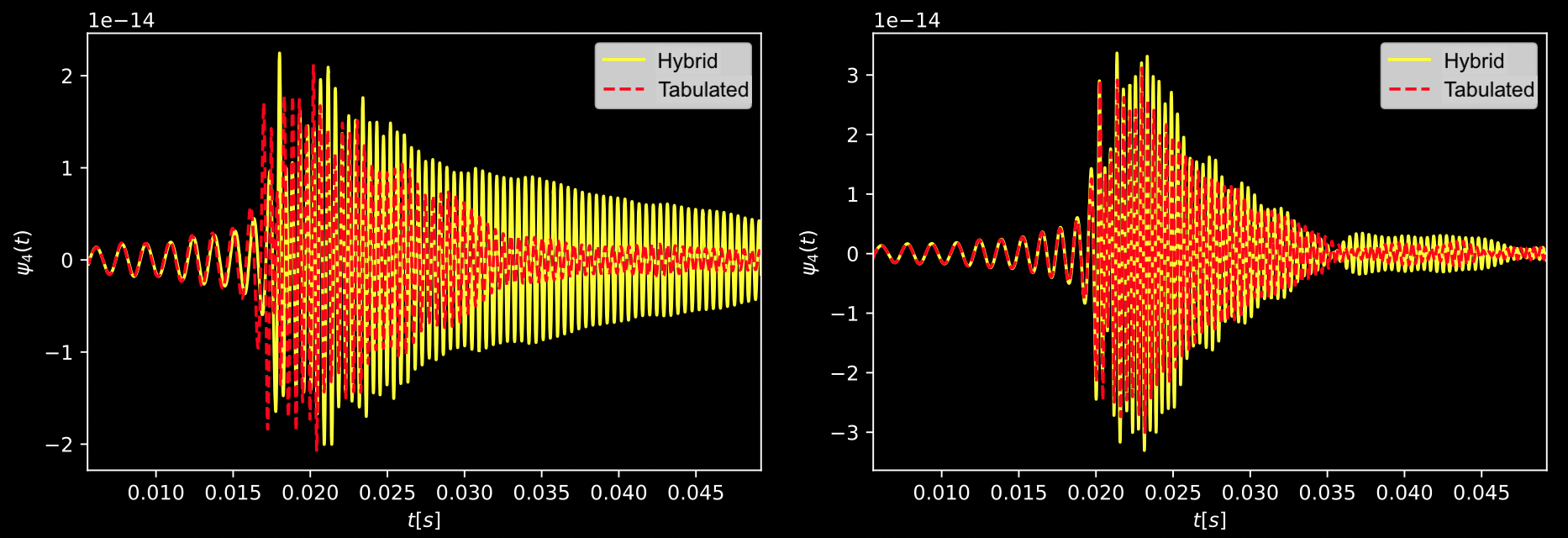}
\caption{Quadrupolar GW signals of two BNS mergers of our sample, \texttt{HShen} EoS (left) and \texttt{DD2} EoS (right). Yellow and red waveforms refer to hybrid and tabulated models, respectively.}
\label{fig:waves}
\end{center}
\end{figure*}

Evolutions were performed using the publicly available {\tt IllinoisGRMHD} code~\citep{Werneck:2022exo,Etienne:2015cea} embedded on the {\tt Einstein Toolkit} infrastructure~\citep{Loffler:2011ay}. The code evolves the Baumgarte–Shapiro–Shibata–Nakamura equations for the spacetime fields~\citep{Baumgarte:1998,Shibata:1995} coupled to the puncture gauge conditions setting the damping coefficient appearing in the shift condition to $1/M_{\rm ADM}$. The {\tt illinoisGRMHD} code adopts the Valencia formalism for the general relativistic hydrodynamics equations~\citep{Valencia} which are integrated numerically with a state-of-the-art finite-volume algorithm. 
The simulations used fourth-order spatial stencils and a fourth-order Runge-Kutta scheme for time integration with a Courant-Friedrichs-Lewy factor of 0.5.
 {Finally, the numerical grid hierarchy consists of three sets of nested refinement boxes, one centered at the ``center of mass" of the binary, and the others two centered on each star. Each 
of them contains five boxes that differ in size and in resolution by factors of two. The finest 
box around the NS has a side half-length of $\sim 1.2\,R_{\rm NS}$, where $R_{\rm NS}$ is the initial NS equatorial radius. This choice allows us to initially  resolve the equatorial NS 
radius by $45$ grid points, what corresponds to a resolution of $\Delta x\sim 244\,\rm m$. For the SLy4 EoS, we also employed a higher resolution resolving the equatorial NS 
radius by $67$ grid points, with a $\Delta x\sim 167\,\rm m$~\cite{Guerra:2025}.}
%
%
\section{Analysis setup}
To assess the observational importance of an accurate implementation of thermal effects in BNS mergers
we perform parameter inference and model selection on numerically simulated GW signals $h(\theta_{\rm true})$, with source parameters $\theta_{\rm true}$. These signals correspond to the very late inspiral, merger and post-merger emission of the BNS models of Table~\ref{tab:initial_parameters}.
We inject them in an idealised three-detector network composed of three Cosmic Explorers \citep{CE,CE2} placed at the locations of LIGO Hanford, LIGO Livingston and Virgo. We perform Bayesian parameter inference on these signals using numerically simulated templates  $h_{\cal{M}}(\theta)$ according to two different emission models  ${\cal{M}}$, respectively including the two alternative implementations of thermal effects (i.e.~hybrid and tabulated).  

The posterior Bayesian probability for source parameters $\theta$ according to a waveform model ${\cal{M}}$ is given by
\begin{equation}
 p_{\cal{M}}(\theta|\theta_{\rm true})=\frac{\pi(\theta){\cal{L}}(\theta|h_{{\cal{M}}}(\theta_{\rm true}))}{{\cal{Z}}_{{\cal{M}}}(\theta|\theta_{\rm true})}.   
\end{equation}
Here, ${\cal{L}}(\theta|h(\theta_{\rm true}))$ denotes the likelihood for the parameters $\theta$ while $\pi(\theta)$ denotes their prior probability. Finally, the term ${\cal{Z}}_{\cal{M}}$ denotes the Bayesian evidence for the model ${\cal{M}}$, given by
\begin{equation}
{\cal{Z}}_{\cal{M}} =\int \pi(\theta){\cal{L}}(\theta|h_{{\cal{M}}}(\theta_{\rm true})) d\theta.  
\end{equation}
For two competing waveform models, their relative Bayes Factor is simply given by 
\begin{equation}
{\cal{B}}^{{\cal{M}}_1}_{{\cal{M}}_2} = 
\frac{{\cal{Z}}_{{\cal{M}}_1}}{{\cal{Z}}_{{\cal{M}}_2}}.
\end{equation}
It is commonly considered that ${\cal{M}}_1$ is strongly preferred with respect to ${{\cal{M}}_2}$ when $\log {\cal{B}}^{{\cal{M}}_1}_{{\cal{M}}_2} > 5$.
%
%
\begin{figure}[t!]
\begin{center}
\includegraphics[width=0.5\textwidth]{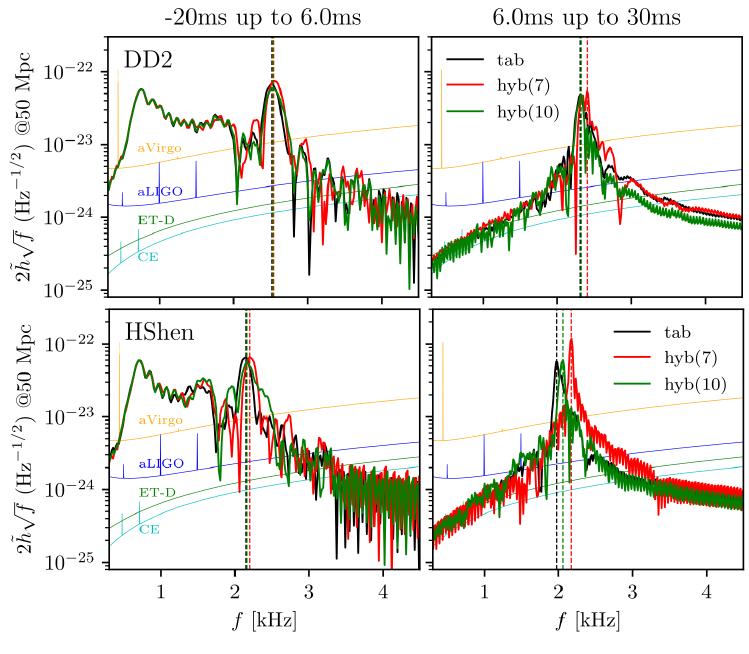}
\caption{ {GW spectra of an optimally oriented BNS merger at 50 Mpc for {\tt DD2} (top) and {\tt HShen} (bottom) EoSs. Black 
lines represent the tabulated  cases, while red and green ones indicates the hybrid cases using seven and ten pieces, respectively
(see Appendix~\ref{Appendix}). 
To highlight the contribution of prominent spectral components at different times,  spectra are shown using two time windows. 
For reference, the design sensitivities of aLIGO~\cite{LIGOsens}, aVirgo~\cite{VIRGO:2014yos}, the Einstein Telescope~\cite{2011CQGra..28i4013H}, 
and Cosmic Explorer~\cite{2017CQGra..34d4001A} are also displayed.}}
\label{fig:spectrum}
\end{center}
\end{figure}

We use the standard frequency-domain likelihood for GW transients  \citep{Finn1992,Romano2017}
\begin{equation}
    \log{\cal{L}}(\theta|h(\theta_{\rm true})) \propto -\sum_N(h(\theta_{\rm true})-h(\theta)|h(\theta_{\rm true})-h(\theta)),
\end{equation}
where $N$ runs over the different detectors of our network. As usual, $(a|b)$ represents the inner product \citep{Cutler1994}
\begin{equation}
   (a|b)= 4 \Re \int_{f_{min}}^{f_{max}} \frac{\tilde{a}(f)\tilde{b}^{*}(f)}{S_n(f)}df,
\end{equation}
where $\tilde{a}(f)$ denotes the Fourier transform of $a(t)$ and ${}^*$ the complex conjugate. The factor $S_{n}(f)$ is the one-sided power spectral density of the detector. We use the predicted Cosmic Explorer sensitivity~\citep{ExploringSens} with a lower frequency cutoff of $f_{\rm min}=600$ Hz and a sampling frequency of $16$ kHz so that $f_{\rm max}=8$ kHz.

Our analysis only includes the merger and post-merger emission, together with the very late inspiral of the process, which is where thermal effects will be most impactful. We note that this part of the signal provides little information about the masses and spins of the binary. In addition, given our usage of numerical relativity simulations, we cannot sample over the individual masses and spins, but only over the extrinsic source parameters, namely the source sky-location, orientation, luminosity distance, polarization angle and time-of-arrival. Nevertheless, it is sensible to assume that masses and spins would be accurately measured from the minutes-long inspiral signal~\citep{Smith2021_BNS,Branchesi:2023mws}. While the same is true for the extrinsic parameters, we choose to actually sample over these, so that we obtain ``conservatively pessimistic'' results that actually underestimate the importance of a ``tabulated'' treatment of thermal effects. We assume standard prior probabilities for the  sky-location, source orientation and polarisation angle, together with a distance prior uniform in co-moving distance and a uniform prior on the time-of-arrival, with a width of $0.2\rm \ s$, centered on the true value. We sample the likelihood across the parameter space using a custom version of the publicly available software \texttt{Parallel Bilby}~\citep{Ashton:2018jfp,pbilby}, sampling the parameters with the algorithm \texttt{Dynesty}~\citep{Dynesty}.\\ 

Finally, we note we do not perform our analysis under the  strain formalism using $h(t)$. Instead, we perform injection and recovery of the Newman-Penrose scalar $\psi_4 = {d^2 h(t)}/{dt^2}$ directly outputted by numerical simulations, using the formalism described in \cite{Psi4PE}. This is done to avoid potential systematic errors arising during the computation of the GW strain $h$ from $\psi_4$ \citep{Pollney_Reissweig}.
%
\section{Results}
\subsection{Merger dynamics and waveforms}
The initial data for the BNS reported in Table~\ref{tab:initial_parameters} were evolved in~\cite{Guerra:2025} for over~$t-t_{\rm mer}\sim 140\,\rm ms$ after merger.\footnote{{Notice that in this section we describe the results obtained from both the tabulated and the hybrid approaches. The latter uses a polytropic representation of the realistic tabulated EoS employing  seven pieces (see Sec.~\ref{sec:Numerical_set}). To assess the robustness of our results against the polytropic representation, 
we compare them in Appendix~\ref{Appendix} with those obtained using ten pieces.}}
Each panel in Fig.~\ref{fig:evo_DD2} displays two snapshots in the evolution of the rest-mass density (left half) and  the thermal specific energy (right half) for both, tabulated (top half) and hybrid cases (bottom half). The top two panels correspond to the \texttt{DD2} EoS and the bottom ones to the \texttt{HShen} EoS. As discussed in~\cite{Guerra:2025}, we use the thermal specific energy $\epsilon_{\rm th}$ as a proxy for the temperature. Only at low densities/high temperatures, the definition of temperature usually employed in BNS merger evolutions based on piecewise polytropes, $T = (\Gamma_{\rm th}-1)\,\epsilon_{\rm th}$ (see e.g.~\cite{DePietri:2019mti}) is equivalent to that of microphysical, finite-temperature  EOS evolutions. Fig.~\ref{fig:evo_DD2} shows that for tabulated evolutions the BNS remnant is hotter near its surface, defined by the density isocontour where $\rho=10^{-3}\rho_{\rm 0, max}$, with $\rho_{\rm 0, max}$ being the initial maximum value of the rest-mass density. This induces a  slightly more extended remnant than that of the hybrid evolutions~\citep{Guerra:2025}.

Fig.~\ref{fig:waves} displays the dominant $\ell=m=2$ GW quadrupole mode of the late inspiral, merger and post-merger emission 
of the simulations corresponding to the same two EoSs of Fig.~\ref{fig:evo_DD2}. The yellow (red) curves indicate hybrid (tabulated) 
evolutions. Differences between the two EoSs are noticeable. The effects of different thermal treatments are clearly more visible 
in the post-merger signal emitted by the \texttt{HShen} case (left panel) for which the amplitude, associated with the oscillations
of the remnant, decays significantly faster. Moreover, small differences between hybrid and tabulated evolutions are noticeable in 
the last cycles of the inspiral signals also in the case of the \texttt{HShen} EoS. This suggests that for this EoS thermal effects may alter the 
tidal deformability in a different  way for both treatments, possibly leading to some observational difference.  {Finally, Fig.~\ref{fig:spectrum} shows the corresponding 
GW spectra assuming an optimally oriented source at a distance of $50$ Mpc. We observe that the hybrid approach induces a frequency shift in the main GW peaks of $\lesssim 300\,\rm Hz$ with respect to the tabulated one~\cite{Guerra:2025} which is significantly reduced when we include 10 pieces.}   
%

%
\begin{table}
    \centering
    \begin{center}
       \caption{{Impact of thermal-effects implementation in gravitational-wave observations} We show the natural log Bayes Factors between tabulated and hybrid models, obtained when these are used to recover a signal from a BNS post-merger using the tabulated EoS.
        We show results for source inclinations of $\iota = \{0.3,0.8,\pi/2\}$ and respective distances of $d_L = \{100,50,10\}$Mpc. We assume a three-detector network composed by three Cosmic Explorers.}
        \begin{tabular}{c|ccc}
            \rule{0pt}{3ex}%
            EoS  &  \multicolumn{3}{c}{$\log {\cal B}^{\text{T}}_{\text{H}}$ ($d_L$ [Mpc], $\rho_{\rm opt}^{\rm net}$)} \\
            \hline
            \rule{0pt}{3ex}%
                            &   $\iota = 0.3$  & $\iota = 0.8$ & $\iota = \pi/2$ \\
                            \rule{0pt}{3ex}%
            \texttt{SLy4}  &  10.1 (100, 15.3)  & 22.1 (50, 22.3) & 38.6 (10, 27.9) \\
            \rule{0pt}{3ex}%
            \texttt{LS220}   &  9.7 (100, 14.3) & 22.1 (50, 20.9)&  33.7 (10, 26.1) \\
            \rule{0pt}{3ex}%
            \texttt{HShen}  &   16.2 (100, 15.1 ) & 37.0 (50, 22.1 ) &  65.2 (10, 27.4 )  \\
            \rule{0pt}{3ex}%
            \texttt{DD2}   &  5.6 (100, 15.3)  & 11.7 (50, 22.4) &  14.3 (10, 27.5) \\
        \end{tabular}
        \label{tab:logb}
    \end{center}
\end{table}
%
%
\subsection{Target binary neutron star sources}
We choose target sources corresponding to four equal-mass BNSs with total mass $M\approx 2.56 M_\odot$ (see Table~\ref{tab:initial_parameters}) characterized by four different EoSs. For each source, we consider three orbital inclinations, namely $\iota=0.3$, $\iota=0.8$ and $\iota=\pi/2$.  Varying the orbital inclination changes the contribution of higher-order harmonics to the observed waveform, which may influence the impact of the thermal effects. While higher-order harmonics are highly suppressed for equal-mass systems, it has been shown that these can be triggered by the so-called ``one-arm'' spiral instability \citep{East2016,Radice2016,Lehner2016}, facilitating the estimation of the source orientation \citep{CaldernBustillo2021}. In particular, our waveforms include the dominant quadrupole modes, $(\ell,m)=(2,\pm 2)$, together with the $(\ell,m)=(2,\pm 1),(3,\pm 3),(3,\pm 2)$ modes.

Ideally, we would perform a wide injection campaign considering sources placed at a large variety of sky-locations, distances and source inclinations. However, the high computational cost of our parameter estimation runs, which make use of numerical relativity, prevents this. For this reason, we only use one fiducial sky-location for our runs. Moreover, the luminosity distance and sky-location do mainly control the signal loudness, without altering the mode content of the signal. Given this, it is expected that the ratio of the maximum signal-to-noise ratios (SNRs) $\rho^{\rm net}$ across the detector network (and therefore the ratio of the respective maximum likelihoods) obtained by the tabulated and hybrid models, should show a weak dependence on these parameters. Since the likelihood, which is what controls the Bayes Factor ${\cal{B}}$, goes as $(\rho^{\rm net})^2/2$, we expect the relative Bayes Factor for the hybrid and tabulated models to go as $(\rho_{\rm opt})^2/2$. Next, the Bayes Factor for each analysis roughly goes as $\log{\cal{B}} \simeq \log{\cal{L}}_{\rm max} - {\cal{C}}$, where ${\cal{C}}$ accounts for the Occam penalty paid by the model. Finally, $\log{\cal{L}}_{\rm max} \simeq {\cal{M}}\times (\rho^{\rm net}_{\rm opt})^2/2$, with ${\cal{M}}$ denoting the match between our injection and the best-fitting template, and $\rho^{\rm net}_{\rm opt}$ indicating the SNR of the injection, equal to the maximal (or optimal) SNR that any analysis can recover~\citep{Lange:2017wki}.

In this situation, for each inclination we perform parameter inference for three fiducial combinations of sky-location and distances yielding reasonably different $\rho_{\rm opt}$. With this, we perform a linear fit $\log{\cal{B}}=\alpha \rho_{\rm max}^2 + \beta$. Next, for each of our target sources, we compute the average SNR over sky-locations at a fiducial distance $d_L^{\rm ref}$. Finally, using that the SNR is inversely proportional to the distance, we compute the distance $d_L^{\rm det}$ at which the averaged relative Bayes Factor between the tabulated and hybrid models satisfies $\log{\cal{B}}^{\rm T}_{\rm H} = 5$, which we will call ``detection'' distance.\\

\subsection{Model selection}

Table \ref{tab:logb} shows $\log{\cal B}^{\rm T}_{\rm H}$ for selected fiducial runs corresponding to each of our selected source inclinations. The parentheses show the luminosity distance chosen for the injection and the corresponding optimal SNR across the detector network. We find that the \texttt{HShen} EoS is most impacted by the choice of implementation of thermal effects, for all source inclinations. This is not a trivial result, as the impact of thermal-effects treatment in the waveform for a given EoS may strongly depend on the source inclination. This may cause some EoS to produce more observable differences for different inclinations. This is observed for the two next cases that are most impacted by the treatment of thermal-effects, namely the \texttt{SLy4} and \texttt{LS220} EoSs. Both cases return similar values of $\log{\cal B}^{\rm T}_{\rm H}$ for the two lowest inclinations. However, the \texttt{SLy4} EoS is most affected for edge-on systems. Finally, the emission from BNS mergers with \texttt{DD2} EoS is the least influenced.

\begin{figure}
\centering
   \includegraphics[width=1.0\linewidth]{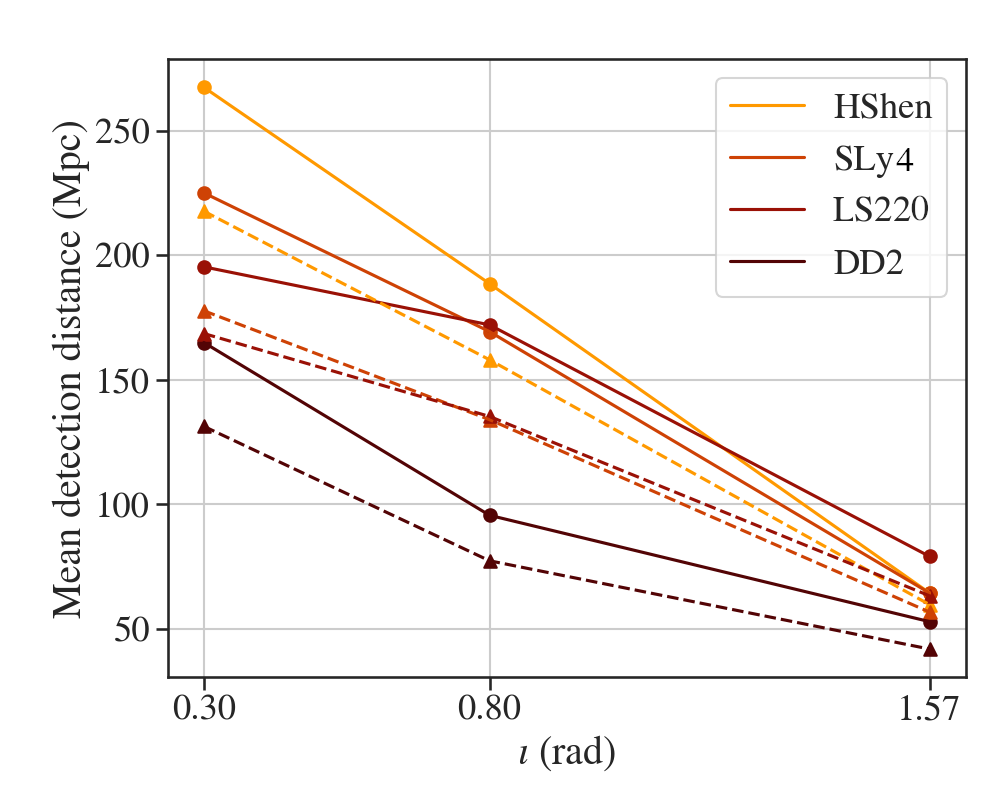}
   \caption{{Impact of thermal effects implementation averaged over source sky-location and azimuthal angle.} Average source distance for which the difference between the log-evidences obtained by the ``tabulated'' and ``hybrid'' models reaches    $\log{\cal{B}}^{\rm T}_{\rm H}=5$, when recoverying a signal modeled through the ``tabulated'' implementation of thermal effects (solid lines). Averaging is performed over source sky-locations and observer's azimuthal angle. Distances at which $\log{\cal{B}}^{\rm T}_{\rm H}=8$ are shown in dashed lines. We show results for and different EoSs and varying source inclinations.}
\label{fig:Detectability}
\end{figure}

Fig.~\ref{fig:Detectability} shows the average ``detection'' distance $d_L^{\rm det}$,
as a function of EoS and source inclination, obtained after performing the fit described in the previous section. We  have  checked empirically that typical deviations from such linear fit are of order $\Delta\log{\cal{B}} \simeq 1 $. To add conservative results we also show lines corresponding to $\log{\cal B}^{\rm T}_{\rm H}=8$. We find that for all EoS, the signal features coming from the ``tabulated'' implementation of thermal effects can be detected at distances $d_L \leq 50$ Mpc regardless of the source inclination, and at distances $d_L = 100$ Mpc for inclinations $\iota < 0.8$. In other words, the correct analysis of a source like GW170817 would require of the more realistic tabulated modeling of its EoS, were the source consistent with one of the EoS of our sample and assuming that other effects our simulations do not yet incorporate, like magnetic fields, bulk and shear viscosity, and neutrinos, do not play a major role\footnote{It should be noted that the accurate numerical modelling of the actual GW spectrum of BNS post-merger remnants is still a matter of debate. In particular, the development of MHD-driven turbulence in the remnant, triggered by the magneto-rotational instability, is likely to affect the evolution of the system, influencing the emission of high-frequency GWs (see e.g.~\cite{Shibata-viscous,Ruiz:2021qmm,Chabanov:2023blf} for recent studies).}. Moreover, the same is true even for weak edge-on sources, for averaged distances of order $50$ Mpc.

To close this section, we note that the use of a simplified hybrid thermal treatment
leads, in certain cases, to significant parameter biases in the sampled parameters, specially the luminosity distance. The biases are due to the different signal amplitudes predicted by tabulated and hybrid models, as it is clear in Fig. \ref{fig:waves}, specially in the case of the \texttt{HShen} EoS. Nevertheless, we stress again that these parameters are expected to be measured with great accuracy from the minute-duration inspiral signal.

%
\section{\textbf{Conclusions}}
Using a combination of a Bayesian framework and 
numerical-relativity simulations, we have shown the dramatic importance of an accurate treatment of thermal effects in the post-merger GW emission of BNS mergers. We have found that, for all EoSs and source inclinations considered, self-consistent ``tabulated'' implementations lead to modifications of the post-merger GW emission with respect to simplified ``hybrid'' approaches that are 
observable with a detector network formed by three Cosmic Explorers, for sources at distances $d_L \leq 50$ Mpc. 
In particular, for inclinations $\iota \leq 0.8$, consistent with GW170817, the differences are visible for distances $d_L \leq 150$ Mpc, with the exception of the \texttt{DD2} EoS. Out of the four EoSs considered and within the limitations of our simulations, the post-merger GW signal of the \texttt{HShen} EoS is the most influenced by a  hybrid implementation of thermal effects.
{To probe the robustness of our results we used two hybrid implementations, using a piecewise regression as in~\cite{2021JOSS....6.3859P} with seven or ten polytropic 
pieces~\citep{Read:2008iy}. While the corresponding waveforms may show significant mismatches compared to those using seven pieces, our qualitative findings remain 
unchanged (see Appendix~\ref{Appendix}).}

We note that our results rather underestimate the importance of a self-consistent implementation of thermal effects, as we perform inference on parameters that are expected to be accurately constrained from the long inspiral signal, like sky-location or source orientation. This allows hybrid models to exploit parameter degeneracies to achieve higher Bayesian evidences than they would do otherwise, if such parameters were accurately constrained through the inspiral signal.

Our work can be regarded as a ``proof of principle'' of the application of parameter inference  to BNS merger remnants using numerical-relativity waveforms, 
which is obviously limited by the fact that we do not sample over the actual EoS. A natural extension of our work would be to compare injections including thermal 
effects for a given EoS with a large set (ideally continuous) of templates covering a wide range of EoSs (as somewhat done in~\cite{Bustillo2021,Psi4PE} for the case 
of collisions of a different matter source, Proca-stars\footnote{Note that, in the Proca-star case, the simulations do not span a continuous range of EoSs, as for 
neutron stars. Instead, they span a very dense grid in the individual frequencies of the Proca fields. }).  {However, the lack of available thermal EoSs, along with
the high cost of our numerical simulations,} together with the intrinsic discreteness and low number of simulations including thermal effects prevents such extension 
 {with the finite computational resources at our disposal}.
Nevertheless, our study proves that waveforms sourced by the EoSs of our sample are sufficiently different so as to allow the identification of the underlying EoS 
for the range of signal loudness we consider.

{A few caveats remain. First, as shown in~\cite{Miravet-Tenes:2023}, the initial data for the two types of BNS models employed in this paper are slightly different. Although they are built as similar as possible to minimize the impact of differences on the simulations, there are intrinsic differences in the way the models are built. In particular, 
the lowest value of the temperature in the   {publicly} available tables is $T = 0.01\,\rm MeV$, which affects the density distribution at low densities, making impossible to build the exact same configurations. Nevertheless, the gravitational mass of the tabulated and hybrid inital configurations differ only by $< 0.2\%$, their circunferencial radius by $< 2\%$, and their tidal deformability number by $< 5\%$. These values are  similar to the intrinsic error in typical numerical simulations of BNS mergers (see e.g.~\cite{Etienne:2011ea,Ruiz:2020elr,Ruiz:2020via,Baiotti:2008ra}) and hence cannot explain the differences reported here.
  Second, recent numerical approaches have attempted to mimic the behaviour of the polytropic constant index $\Gamma_{\rm th}$ above nuclear saturation density~(see  e.g.~\cite{Raithel:2023zml}), which may potentially improve the way thermal effects are handled in the hybrid approach. However, it remains unclear if this approach is accurate enough to capture fully the thermal effects of a tabulated EoS. Further studies are necessary to determine if the former approach induced negligible differences in the post-merger GW that could not be observed by third-generation GW detectors. It is worth mentioning that several general relativistic simulations of binary neutron star mergers~(see~e.g.~\cite{Bauswein:2010dn,Rivieccio:2024sfm,Takami:2014tva}) evolved using the hybrid approach have shown that changes in $\Gamma_{\rm th}$ ranging between $\sim 1.7$ and $2$ induce  changes in the frequency of the main GW modes of $\lesssim 8\%$. Therefore, varying $\Gamma_{\rm th}$ within this range is unlikely to lead to significantly different outcomes. Conducting a dedicated survey of $\Gamma_{\rm th}$ for different EoSs is 
  beyond the scope.}  {Third, some EoSs account for composition changes, such as hyperon production, which softens the EoS by reducing pressure as highly degenerate nucleons 
  are depopulated. Recently, Blacker et al.~\cite{Blacker:2023opp} showed that this phenomenon may shift the frequency of the main GW peak in $\sim 150\,\rm Hz$ compared 
  to purely nucleonic EoSs, which can be potentially detected if the EoS and stellar parameters of cold neutron stars are known. Notice that these studies do not include
  magnetic viscosity. Recently, it was shown that magnetic viscosity may also induce frequency shifts (see~\cite{Bamber:2024qzi,Tsokaros:2024wgb}). Nevertheless, these 
  changes in the frequency are typically smaller than those we reported between the hybrid and tabulated treatments of the thermal effects.}

\appendix
\section{Making use of 10-piece polytrope waveforms}
\label{Appendix}

{In the main text we compared tabulated waveforms to  hybrid waveforms where the cold part
of the EoS is modeled using seven polytropic pieces. While such modeling represents the current state-of-the art, it is reasonable 
to ask whether our results would hold if more complete representations making use of a larger number of polytropic pieces were used.  
To this end, in this appendix we repeat our study with new hybrid simulations consisting of ten pieces.
It is worth noting that increasing the number of pieces makes the numerical simulation more computationally expensive. For this reason,
most BNS simulations typically use no more than seven pieces~(see~e.g.~\cite{Rivieccio:2024sfm,Bamber:2024kfb}).}

\begin{figure}
\centering
   \includegraphics[width=1.0\linewidth]{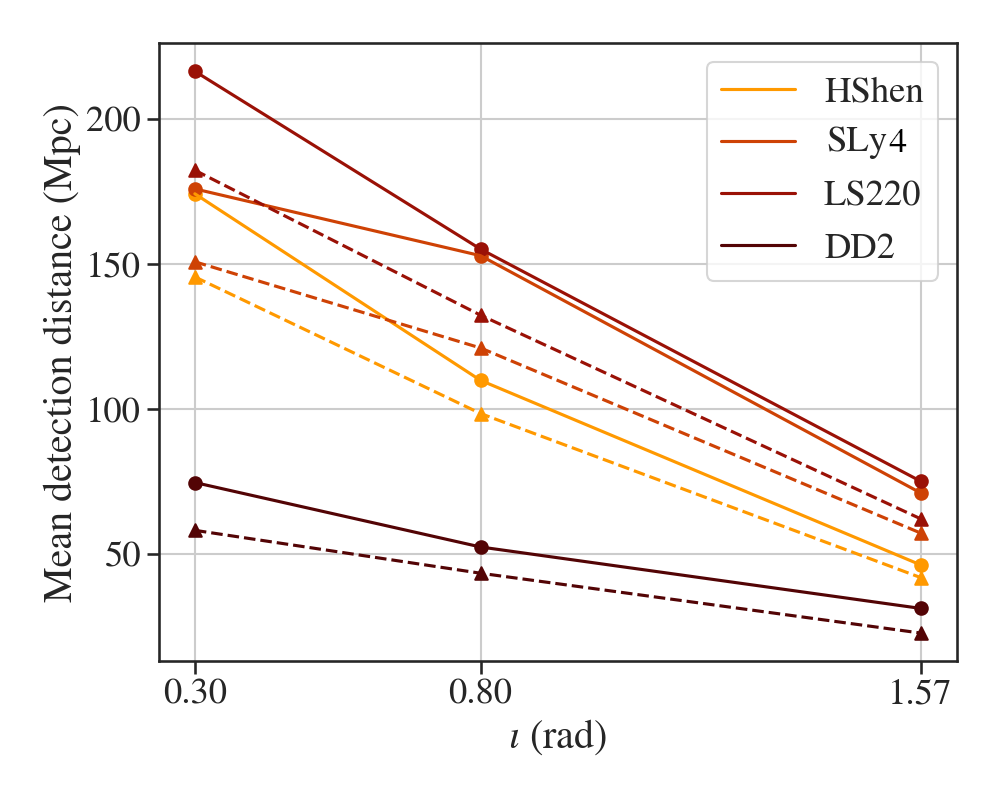}
   \caption{{Impact of thermal effects implementation, as a function of the source inclination, averaged over source sky-location and azimuthal 
   angle, when hybrid post-merger waveforms are modelled with ten polytropic pieces.} }
\label{fig:Detectability10}
\end{figure}
{Fig.~\ref{fig:Detectability10} shows the same results as Fig.~\ref{fig:Detectability} in the main text, with the difference that ten pieces are used to model the hybrid EoS. The variation in the detectability of the treatment of thermal effects can be roughly determined by the similarity between our new hybrid waveforms and the tabulated ones. The similarity between two waveforms $a(t)$ and $b(t)$ can be quantified through the so-called match, defined as the overlap ${\cal{O}}(a|b) = (a|b)/\sqrt{(a|a)(b|b)}$ optimised over relative time and phase shifts. We show in Table~\ref{tab:overlaps} the match between the dominant $(\ell,m)=(2,2)$ modes of our tabulated simulations and our hybrid simulations with 7 and 10 pieces. We observe that the match is significantly improved for the \texttt{HShen} and \texttt{DD2} EoS (it gets closer to one) while it is slightly reduced for the other two}.\\

{The above matches result in a reduction of the distances needed to resolve the modelling of thermal effects for low inclinations (for which the observed signal is vastly dominated by the $(2,2)$ mode) for the \texttt{HShen} and \texttt{DD2} EoS. Conversely, such distance is increased for the \texttt{LS220} case, which is now the EoS for which the treatment of thermal effects is easier to resolve. In this case the differences can be told apart at distances as large as about $220$ Mpc, as compared to the $180$ Mpc obtained for 7-piece polytropes.  Also, the detection distance is greatly reduced for the \texttt{DD2} case, from $170$ Mpc and $100$ Mpc for, respectively, low and mid inclinations to $75$ Mpc and $50$ Mpc. This reflects the fact that the match between  hybrid and tabulated waveforms increases from $0.97$ to $0.99$ for this case. We also observe that the detection distance for the \texttt{SLy} EoS is reduced despite the reduced match with the tabulated waveforms. This, however, can be simply due to the influence of extra gravitational-wave modes, which may translate to net waveforms less similar to the tabulated ones, or to an increased Occam penalty for the hybrid model.} \\

{Overall, we find that our qualitative conclusions remain largely unaffected by the use of 10-piece waveforms. In particular, in all cases the differences in the treatment of thermal effects are detectable at around $\sim 50$ Mpc and only the \texttt{DD2} EoS prevents their detectability at distances of $\sim 150$ Mpc.} \\
\begin{table}
    \centering   
    \begin{center}
        \begin{tabular}{c|cc}
              \texttt{EoS}  &  7-piece & 10-piece \\
            \hline
            \rule{0pt}{3ex}%
            \texttt{SLy4}  &  0.900 & 0.888 \\
            \rule{0pt}{3ex}%
            \texttt{LS220}   &  0.916 &  0.894 \\
            \rule{0pt}{3ex}%
            \texttt{HShen}  &   0.899 &  0.944 \\
            \rule{0pt}{3ex}%
            \texttt{DD2}   &  0.966 & 0.991 \\
        \end{tabular}
        \label{tab:overlaps}
          \caption{{Similarity between tabulated and hybrid waveforms} Matches between the dominant $(\ell,m)=(2,2)$ modes of our tabulated waveforms and those of our hybrid waveforms 
     computed with seven and ten polytropic pieces.}
    \end{center}
\end{table}
%
\acknowledgments
It is a pleasure to thank Zachariah Etienne and Leonardo Werneck for  useful discussions.
This work was supported by a fellowship from ``la Caixa'' Foundation (ID100010434), the European Union’s Horizon2020 research and innovation programme under the Marie Skłodowska-Curie grant agreement No 847648 (fellowship code LCF/BQ/PI20/11760016), the Generalitat Valenciana through the grants CIDEGENT/2021/046
and Prometeo CIPROM/2022/49, and the Spanish Agencia Estatal de  Investigaci\'on through the grants PRE2019-087617, PID2020-118635GB-I00 and PID2021-125485NB-C21 funded by MCIN/AEI/10.13039/501100011033 and ERDF A way of making Europe. Further support has been provided by the EU's Horizon 2020 Research and Innovation (RISE) programme H2020-MSCA-RISE-2017 (FunFiCO-777740) and  by  the  EU  Staff  Exchange  (SE)  programme HORIZON-MSCA-2021-SE-01 (NewFunFiCO-101086251).
The authors acknowledge computational
resources provided by the LIGO Laboratory and supported by
National Science Foundation Grants PHY-0757058 and PHY0823459, the support of the NSF CIT cluster for the provision of computational resources for our parameter inference runs, and the computational resources and technical support of the Spanish Supercomputing Network through the use of MareNostrum at the Barcelona Supercomputing Center (AECT-2023-1-0006) where the BNS merger simulations were performed.
This material is based upon work supported by NSF's LIGO Laboratory which is a major facility fully funded by the National Science Foundation. This document has LIGO DCC number LIGO-P2300344.

%
\bibliographystyle{apsrev4-1}
\bibliography{ref}

\begin{thebibliography}{94}%
\makeatletter
\providecommand \@ifxundefined [1]{%
 \@ifx{#1\undefined}
}%
\providecommand \@ifnum [1]{%
 \ifnum #1\expandafter \@firstoftwo
 \else \expandafter \@secondoftwo
 \fi
}%
\providecommand \@ifx [1]{%
 \ifx #1\expandafter \@firstoftwo
 \else \expandafter \@secondoftwo
 \fi
}%
\providecommand \natexlab [1]{#1}%
\providecommand \enquote  [1]{``#1''}%
\providecommand \bibnamefont  [1]{#1}%
\providecommand \bibfnamefont [1]{#1}%
\providecommand \citenamefont [1]{#1}%
\providecommand \href@noop [0]{\@secondoftwo}%
\providecommand \href [0]{\begingroup \@sanitize@url \@href}%
\providecommand \@href[1]{\@@startlink{#1}\@@href}%
\providecommand \@@href[1]{\endgroup#1\@@endlink}%
\providecommand \@sanitize@url [0]{\catcode `\\12\catcode `\$12\catcode
  `\&12\catcode `\#12\catcode `\^12\catcode `\_12\catcode `\%12\relax}%
\providecommand \@@startlink[1]{}%
\providecommand \@@endlink[0]{}%
\providecommand \url  [0]{\begingroup\@sanitize@url \@url }%
\providecommand \@url [1]{\endgroup\@href {#1}{\urlprefix }}%
\providecommand \urlprefix  [0]{URL }%
\providecommand \Eprint [0]{\href }%
\providecommand \doibase [0]{http://dx.doi.org/}%
\providecommand \selectlanguage [0]{\@gobble}%
\providecommand \bibinfo  [0]{\@secondoftwo}%
\providecommand \bibfield  [0]{\@secondoftwo}%
\providecommand \translation [1]{[#1]}%
\providecommand \BibitemOpen [0]{}%
\providecommand \bibitemStop [0]{}%
\providecommand \bibitemNoStop [0]{.\EOS\space}%
\providecommand \EOS [0]{\spacefactor3000\relax}%
\providecommand \BibitemShut  [1]{\csname bibitem#1\endcsname}%
\let\auto@bib@innerbib\@empty
\bibitem [{\citenamefont {Abbott}\ \emph {et~al.}(2021)\citenamefont {Abbott}
  \emph {et~al.}}]{LIGOScientific:2021djp}%
  \BibitemOpen
  \bibfield  {author} {\bibinfo {author} {\bibfnamefont {R.}~\bibnamefont
  {Abbott}} \emph {et~al.} (\bibinfo {collaboration} {LIGO Scientific, VIRGO,
  KAGRA}),\ }\href@noop {} {\  (\bibinfo {year} {2021})},\ \Eprint
  {http://arxiv.org/abs/2111.03606} {arXiv:2111.03606 [gr-qc]} \BibitemShut
  {NoStop}%
\bibitem [{\citenamefont {Abbott}\ \emph
  {et~al.}(2017{\natexlab{a}})\citenamefont {Abbott} \emph
  {et~al.}}]{GBM:2017lvd}%
  \BibitemOpen
  \bibfield  {author} {\bibinfo {author} {\bibfnamefont {B.~P.}\ \bibnamefont
  {Abbott}} \emph {et~al.},\ }\href {\doibase 10.3847/2041-8213/aa91c9}
  {\bibfield  {journal} {\bibinfo  {journal} {Astrophys. J.}\ }\textbf
  {\bibinfo {volume} {848}},\ \bibinfo {pages} {L12} (\bibinfo {year}
  {2017}{\natexlab{a}})},\ \Eprint {http://arxiv.org/abs/1710.05833}
  {arXiv:1710.05833 [astro-ph.HE]} \BibitemShut {NoStop}%
\bibitem [{\citenamefont {Abbott}\ \emph
  {et~al.}(2017{\natexlab{b}})\citenamefont {Abbott} \emph
  {et~al.}}]{LIGOScientific:2017zic}%
  \BibitemOpen
  \bibfield  {author} {\bibinfo {author} {\bibfnamefont {B.~P.}\ \bibnamefont
  {Abbott}} \emph {et~al.} (\bibinfo {collaboration} {LIGO Scientific, Virgo,
  Fermi-GBM, INTEGRAL}),\ }\href {\doibase 10.3847/2041-8213/aa920c} {\bibfield
   {journal} {\bibinfo  {journal} {Astrophys. J. Lett.}\ }\textbf {\bibinfo
  {volume} {848}},\ \bibinfo {pages} {L13} (\bibinfo {year}
  {2017}{\natexlab{b}})},\ \Eprint {http://arxiv.org/abs/1710.05834}
  {arXiv:1710.05834 [astro-ph.HE]} \BibitemShut {NoStop}%
\bibitem [{\citenamefont {Abbott}\ \emph
  {et~al.}(2017{\natexlab{c}})\citenamefont {Abbott} \emph
  {et~al.}}]{LIGOScientific:2017pwl}%
  \BibitemOpen
  \bibfield  {author} {\bibinfo {author} {\bibfnamefont {B.~P.}\ \bibnamefont
  {Abbott}} \emph {et~al.} (\bibinfo {collaboration} {LIGO Scientific,
  Virgo}),\ }\href {\doibase 10.3847/2041-8213/aa9478} {\bibfield  {journal}
  {\bibinfo  {journal} {Astrophys. J. Lett.}\ }\textbf {\bibinfo {volume}
  {850}},\ \bibinfo {pages} {L39} (\bibinfo {year} {2017}{\natexlab{c}})},\
  \Eprint {http://arxiv.org/abs/1710.05836} {arXiv:1710.05836 [astro-ph.HE]}
  \BibitemShut {NoStop}%
\bibitem [{\citenamefont {Shibata}\ \emph {et~al.}(2017)\citenamefont
  {Shibata}, \citenamefont {Fujibayashi}, \citenamefont {Hotokezaka},
  \citenamefont {Kiuchi}, \citenamefont {Kyutoku}, \citenamefont {Sekiguchi},\
  and\ \citenamefont {Tanaka}}]{Shibata:2017xdx}%
  \BibitemOpen
  \bibfield  {author} {\bibinfo {author} {\bibfnamefont {M.}~\bibnamefont
  {Shibata}}, \bibinfo {author} {\bibfnamefont {S.}~\bibnamefont
  {Fujibayashi}}, \bibinfo {author} {\bibfnamefont {K.}~\bibnamefont
  {Hotokezaka}}, \bibinfo {author} {\bibfnamefont {K.}~\bibnamefont {Kiuchi}},
  \bibinfo {author} {\bibfnamefont {K.}~\bibnamefont {Kyutoku}}, \bibinfo
  {author} {\bibfnamefont {Y.}~\bibnamefont {Sekiguchi}}, \ and\ \bibinfo
  {author} {\bibfnamefont {M.}~\bibnamefont {Tanaka}},\ }\href {\doibase
  10.1103/PhysRevD.96.123012} {\bibfield  {journal} {\bibinfo  {journal} {Phys.
  Rev. D}\ }\textbf {\bibinfo {volume} {96}},\ \bibinfo {pages} {123012}
  (\bibinfo {year} {2017})},\ \Eprint {http://arxiv.org/abs/1710.07579}
  {arXiv:1710.07579 [astro-ph.HE]} \BibitemShut {NoStop}%
\bibitem [{\citenamefont {Margalit}\ and\ \citenamefont
  {Metzger}(2017)}]{Margalit:2017dij}%
  \BibitemOpen
  \bibfield  {author} {\bibinfo {author} {\bibfnamefont {B.}~\bibnamefont
  {Margalit}}\ and\ \bibinfo {author} {\bibfnamefont {B.~D.}\ \bibnamefont
  {Metzger}},\ }\href {\doibase 10.3847/2041-8213/aa991c} {\bibfield  {journal}
  {\bibinfo  {journal} {Astrophys. J. Lett.}\ }\textbf {\bibinfo {volume}
  {850}},\ \bibinfo {pages} {L19} (\bibinfo {year} {2017})},\ \Eprint
  {http://arxiv.org/abs/1710.05938} {arXiv:1710.05938 [astro-ph.HE]}
  \BibitemShut {NoStop}%
\bibitem [{\citenamefont {Rezzolla}\ \emph {et~al.}(2018)\citenamefont
  {Rezzolla}, \citenamefont {Most},\ and\ \citenamefont
  {Weih}}]{Rezzolla:2017aly}%
  \BibitemOpen
  \bibfield  {author} {\bibinfo {author} {\bibfnamefont {L.}~\bibnamefont
  {Rezzolla}}, \bibinfo {author} {\bibfnamefont {E.~R.}\ \bibnamefont {Most}},
  \ and\ \bibinfo {author} {\bibfnamefont {L.~R.}\ \bibnamefont {Weih}},\
  }\href {\doibase 10.3847/2041-8213/aaa401} {\bibfield  {journal} {\bibinfo
  {journal} {Astrophys. J. Lett.}\ }\textbf {\bibinfo {volume} {852}},\
  \bibinfo {pages} {L25} (\bibinfo {year} {2018})},\ \Eprint
  {http://arxiv.org/abs/1711.00314} {arXiv:1711.00314 [astro-ph.HE]}
  \BibitemShut {NoStop}%
\bibitem [{\citenamefont {Ruiz}\ \emph {et~al.}(2018)\citenamefont {Ruiz},
  \citenamefont {Shapiro},\ and\ \citenamefont {Tsokaros}}]{Ruiz:2017due}%
  \BibitemOpen
  \bibfield  {author} {\bibinfo {author} {\bibfnamefont {M.}~\bibnamefont
  {Ruiz}}, \bibinfo {author} {\bibfnamefont {S.~L.}\ \bibnamefont {Shapiro}}, \
  and\ \bibinfo {author} {\bibfnamefont {A.}~\bibnamefont {Tsokaros}},\ }\href
  {\doibase 10.1103/PhysRevD.97.021501} {\bibfield  {journal} {\bibinfo
  {journal} {Phys. Rev. D}\ }\textbf {\bibinfo {volume} {97}},\ \bibinfo
  {pages} {021501} (\bibinfo {year} {2018})},\ \Eprint
  {http://arxiv.org/abs/1711.00473} {arXiv:1711.00473 [astro-ph.HE]}
  \BibitemShut {NoStop}%
\bibitem [{\citenamefont {Bauswein}\ \emph {et~al.}(2017)\citenamefont
  {Bauswein}, \citenamefont {Just}, \citenamefont {Janka},\ and\ \citenamefont
  {Stergioulas}}]{Bauswein:2017vtn}%
  \BibitemOpen
  \bibfield  {author} {\bibinfo {author} {\bibfnamefont {A.}~\bibnamefont
  {Bauswein}}, \bibinfo {author} {\bibfnamefont {O.}~\bibnamefont {Just}},
  \bibinfo {author} {\bibfnamefont {H.-T.}\ \bibnamefont {Janka}}, \ and\
  \bibinfo {author} {\bibfnamefont {N.}~\bibnamefont {Stergioulas}},\ }\href
  {\doibase 10.3847/2041-8213/aa9994} {\bibfield  {journal} {\bibinfo
  {journal} {Astrophys. J. Lett.}\ }\textbf {\bibinfo {volume} {850}},\
  \bibinfo {pages} {L34} (\bibinfo {year} {2017})},\ \Eprint
  {http://arxiv.org/abs/1710.06843} {arXiv:1710.06843 [astro-ph.HE]}
  \BibitemShut {NoStop}%
\bibitem [{\citenamefont {Most}\ \emph {et~al.}(2018)\citenamefont {Most},
  \citenamefont {Weih}, \citenamefont {Rezzolla},\ and\ \citenamefont
  {Schaffner-Bielich}}]{Most:2018hfd}%
  \BibitemOpen
  \bibfield  {author} {\bibinfo {author} {\bibfnamefont {E.~R.}\ \bibnamefont
  {Most}}, \bibinfo {author} {\bibfnamefont {L.~R.}\ \bibnamefont {Weih}},
  \bibinfo {author} {\bibfnamefont {L.}~\bibnamefont {Rezzolla}}, \ and\
  \bibinfo {author} {\bibfnamefont {J.}~\bibnamefont {Schaffner-Bielich}},\
  }\href {\doibase 10.1103/PhysRevLett.120.261103} {\bibfield  {journal}
  {\bibinfo  {journal} {Phys. Rev. Lett.}\ }\textbf {\bibinfo {volume} {120}},\
  \bibinfo {pages} {261103} (\bibinfo {year} {2018})},\ \Eprint
  {http://arxiv.org/abs/1803.00549} {arXiv:1803.00549 [gr-qc]} \BibitemShut
  {NoStop}%
\bibitem [{\citenamefont {Ezquiaga}\ and\ \citenamefont
  {Zumalac\'arregui}(2017)}]{Ezquiaga:2017ekz}%
  \BibitemOpen
  \bibfield  {author} {\bibinfo {author} {\bibfnamefont {J.~M.}\ \bibnamefont
  {Ezquiaga}}\ and\ \bibinfo {author} {\bibfnamefont {M.}~\bibnamefont
  {Zumalac\'arregui}},\ }\href {\doibase 10.1103/PhysRevLett.119.251304}
  {\bibfield  {journal} {\bibinfo  {journal} {Phys. Rev. Lett.}\ }\textbf
  {\bibinfo {volume} {119}},\ \bibinfo {pages} {251304} (\bibinfo {year}
  {2017})},\ \Eprint {http://arxiv.org/abs/1710.05901} {arXiv:1710.05901
  [astro-ph.CO]} \BibitemShut {NoStop}%
\bibitem [{\citenamefont {Pardo}\ \emph {et~al.}(2018)\citenamefont {Pardo},
  \citenamefont {Fishbach}, \citenamefont {Holz},\ and\ \citenamefont
  {Spergel}}]{Pardo:2018ipy}%
  \BibitemOpen
  \bibfield  {author} {\bibinfo {author} {\bibfnamefont {K.}~\bibnamefont
  {Pardo}}, \bibinfo {author} {\bibfnamefont {M.}~\bibnamefont {Fishbach}},
  \bibinfo {author} {\bibfnamefont {D.~E.}\ \bibnamefont {Holz}}, \ and\
  \bibinfo {author} {\bibfnamefont {D.~N.}\ \bibnamefont {Spergel}},\ }\href
  {\doibase 10.1088/1475-7516/2018/07/048} {\bibfield  {journal} {\bibinfo
  {journal} {JCAP}\ }\textbf {\bibinfo {volume} {07}},\ \bibinfo {pages} {048}
  (\bibinfo {year} {2018})},\ \Eprint {http://arxiv.org/abs/1801.08160}
  {arXiv:1801.08160 [gr-qc]} \BibitemShut {NoStop}%
\bibitem [{\citenamefont {Abbott}\ \emph
  {et~al.}(2017{\natexlab{d}})\citenamefont {Abbott} \emph
  {et~al.}}]{LIGOScientific:2017adf}%
  \BibitemOpen
  \bibfield  {author} {\bibinfo {author} {\bibfnamefont {B.~P.}\ \bibnamefont
  {Abbott}} \emph {et~al.} (\bibinfo {collaboration} {LIGO Scientific, Virgo,
  1M2H, Dark Energy Camera GW-E, DES, DLT40, Las Cumbres Observatory, VINROUGE,
  MASTER}),\ }\href {\doibase 10.1038/nature24471} {\bibfield  {journal}
  {\bibinfo  {journal} {Nature}\ }\textbf {\bibinfo {volume} {551}},\ \bibinfo
  {pages} {85} (\bibinfo {year} {2017}{\natexlab{d}})},\ \Eprint
  {http://arxiv.org/abs/1710.05835} {arXiv:1710.05835 [astro-ph.CO]}
  \BibitemShut {NoStop}%
\bibitem [{\citenamefont {Dietrich}\ \emph {et~al.}(2020)\citenamefont
  {Dietrich}, \citenamefont {Coughlin}, \citenamefont {Pang}, \citenamefont
  {Bulla}, \citenamefont {Heinzel}, \citenamefont {Issa}, \citenamefont
  {Tews},\ and\ \citenamefont {Antier}}]{Dietrich:2020efo}%
  \BibitemOpen
  \bibfield  {author} {\bibinfo {author} {\bibfnamefont {T.}~\bibnamefont
  {Dietrich}}, \bibinfo {author} {\bibfnamefont {M.~W.}\ \bibnamefont
  {Coughlin}}, \bibinfo {author} {\bibfnamefont {P.~T.~H.}\ \bibnamefont
  {Pang}}, \bibinfo {author} {\bibfnamefont {M.}~\bibnamefont {Bulla}},
  \bibinfo {author} {\bibfnamefont {J.}~\bibnamefont {Heinzel}}, \bibinfo
  {author} {\bibfnamefont {L.}~\bibnamefont {Issa}}, \bibinfo {author}
  {\bibfnamefont {I.}~\bibnamefont {Tews}}, \ and\ \bibinfo {author}
  {\bibfnamefont {S.}~\bibnamefont {Antier}},\ }\href {\doibase
  10.1126/science.abb4317} {\bibfield  {journal} {\bibinfo  {journal}
  {Science}\ }\textbf {\bibinfo {volume} {370}},\ \bibinfo {pages} {1450}
  (\bibinfo {year} {2020})},\ \Eprint {http://arxiv.org/abs/2002.11355}
  {arXiv:2002.11355 [astro-ph.HE]} \BibitemShut {NoStop}%
\bibitem [{\citenamefont {{Metzger}}\ \emph {et~al.}(2010)\citenamefont
  {{Metzger}}, \citenamefont {{Mart{\'\i}nez-Pinedo}}, \citenamefont
  {{Darbha}}, \citenamefont {{Quataert}}, \citenamefont {{Arcones}},
  \citenamefont {{Kasen}}, \citenamefont {{Thomas}}, \citenamefont {{Nugent}},
  \citenamefont {{Panov}},\ and\ \citenamefont
  {{Zinner}}}]{2010MNRAS.406.2650M}%
  \BibitemOpen
  \bibfield  {author} {\bibinfo {author} {\bibfnamefont {B.~D.}\ \bibnamefont
  {{Metzger}}}, \bibinfo {author} {\bibfnamefont {G.}~\bibnamefont
  {{Mart{\'\i}nez-Pinedo}}}, \bibinfo {author} {\bibfnamefont {S.}~\bibnamefont
  {{Darbha}}}, \bibinfo {author} {\bibfnamefont {E.}~\bibnamefont
  {{Quataert}}}, \bibinfo {author} {\bibfnamefont {A.}~\bibnamefont
  {{Arcones}}}, \bibinfo {author} {\bibfnamefont {D.}~\bibnamefont {{Kasen}}},
  \bibinfo {author} {\bibfnamefont {R.}~\bibnamefont {{Thomas}}}, \bibinfo
  {author} {\bibfnamefont {P.}~\bibnamefont {{Nugent}}}, \bibinfo {author}
  {\bibfnamefont {I.~V.}\ \bibnamefont {{Panov}}}, \ and\ \bibinfo {author}
  {\bibfnamefont {N.~T.}\ \bibnamefont {{Zinner}}},\ }\href {\doibase
  10.1111/j.1365-2966.2010.16864.x} {\bibfield  {journal} {\bibinfo  {journal}
  {Mon. Not. Roy. Astron. Soc.}\ }\textbf {\bibinfo {volume} {406}},\ \bibinfo
  {pages} {2650} (\bibinfo {year} {2010})},\ \Eprint
  {http://arxiv.org/abs/1001.5029} {arXiv:1001.5029 [astro-ph.HE]} \BibitemShut
  {NoStop}%
\bibitem [{\citenamefont {{Cowperthwaite}}\ \emph {et~al.}(2017)\citenamefont
  {{Cowperthwaite}} \emph {et~al.}}]{Cowperthwaite:2017}%
  \BibitemOpen
  \bibfield  {author} {\bibinfo {author} {\bibfnamefont {P.~S.}\ \bibnamefont
  {{Cowperthwaite}}} \emph {et~al.},\ }\href {\doibase
  10.3847/2041-8213/aa8fc7} {\bibfield  {journal} {\bibinfo  {journal}
  {Astrophys. J. Lett.}\ }\textbf {\bibinfo {volume} {848}},\ \bibinfo {eid}
  {L17} (\bibinfo {year} {2017})},\ \Eprint {http://arxiv.org/abs/1710.05840}
  {arXiv:1710.05840 [astro-ph.HE]} \BibitemShut {NoStop}%
\bibitem [{\citenamefont {{Kasen}}\ \emph {et~al.}(2017)\citenamefont
  {{Kasen}}, \citenamefont {{Metzger}}, \citenamefont {{Barnes}}, \citenamefont
  {{Quataert}},\ and\ \citenamefont {{Ramirez-Ruiz}}}]{Kasen:2017}%
  \BibitemOpen
  \bibfield  {author} {\bibinfo {author} {\bibfnamefont {D.}~\bibnamefont
  {{Kasen}}}, \bibinfo {author} {\bibfnamefont {B.}~\bibnamefont {{Metzger}}},
  \bibinfo {author} {\bibfnamefont {J.}~\bibnamefont {{Barnes}}}, \bibinfo
  {author} {\bibfnamefont {E.}~\bibnamefont {{Quataert}}}, \ and\ \bibinfo
  {author} {\bibfnamefont {E.}~\bibnamefont {{Ramirez-Ruiz}}},\ }\href
  {\doibase 10.1038/nature24453} {\bibfield  {journal} {\bibinfo  {journal}
  {\nat}\ }\textbf {\bibinfo {volume} {551}},\ \bibinfo {pages} {80} (\bibinfo
  {year} {2017})},\ \Eprint {http://arxiv.org/abs/1710.05463} {arXiv:1710.05463
  [astro-ph.HE]} \BibitemShut {NoStop}%
\bibitem [{\citenamefont {{Pian}}\ \emph {et~al.}(2017)\citenamefont {{Pian}}
  \emph {et~al.}}]{Pian:2017}%
  \BibitemOpen
  \bibfield  {author} {\bibinfo {author} {\bibfnamefont {E.}~\bibnamefont
  {{Pian}}} \emph {et~al.},\ }\href {\doibase 10.1038/nature24298} {\bibfield
  {journal} {\bibinfo  {journal} {\nat}\ }\textbf {\bibinfo {volume} {551}},\
  \bibinfo {pages} {67} (\bibinfo {year} {2017})},\ \Eprint
  {http://arxiv.org/abs/1710.05858} {arXiv:1710.05858 [astro-ph.HE]}
  \BibitemShut {NoStop}%
\bibitem [{\citenamefont {Abbott}\ \emph
  {et~al.}(2017{\natexlab{e}})\citenamefont {Abbott} \emph
  {et~al.}}]{GW170817-post-merger}%
  \BibitemOpen
  \bibfield  {author} {\bibinfo {author} {\bibfnamefont {B.~P.}\ \bibnamefont
  {Abbott}} \emph {et~al.} (\bibinfo {collaboration} {LIGO Scientific,
  Virgo}),\ }\href {\doibase 10.3847/2041-8213/aa9a35} {\bibfield  {journal}
  {\bibinfo  {journal} {Astrophys. J. Lett.}\ }\textbf {\bibinfo {volume}
  {851}},\ \bibinfo {pages} {L16} (\bibinfo {year} {2017}{\natexlab{e}})},\
  \Eprint {http://arxiv.org/abs/1710.09320} {arXiv:1710.09320 [astro-ph.HE]}
  \BibitemShut {NoStop}%
\bibitem [{\citenamefont {{Clark}}\ \emph {et~al.}(2016)\citenamefont
  {{Clark}}, \citenamefont {{Bauswein}}, \citenamefont {{Stergioulas}},\ and\
  \citenamefont {{Shoemaker}}}]{Clark:2016}%
  \BibitemOpen
  \bibfield  {author} {\bibinfo {author} {\bibfnamefont {J.~A.}\ \bibnamefont
  {{Clark}}}, \bibinfo {author} {\bibfnamefont {A.}~\bibnamefont {{Bauswein}}},
  \bibinfo {author} {\bibfnamefont {N.}~\bibnamefont {{Stergioulas}}}, \ and\
  \bibinfo {author} {\bibfnamefont {D.}~\bibnamefont {{Shoemaker}}},\ }\href
  {\doibase 10.1088/0264-9381/33/8/085003} {\bibfield  {journal} {\bibinfo
  {journal} {Classical and Quantum Gravity}\ }\textbf {\bibinfo {volume}
  {33}},\ \bibinfo {eid} {085003} (\bibinfo {year} {2016})},\ \Eprint
  {http://arxiv.org/abs/1509.08522} {arXiv:1509.08522 [astro-ph.HE]}
  \BibitemShut {NoStop}%
\bibitem [{\citenamefont {Miravet-Ten\'es}\ \emph {et~al.}(2023)\citenamefont
  {Miravet-Ten\'es}, \citenamefont {Castillo}, \citenamefont {De~Pietri},
  \citenamefont {Cerd\'a-Dur\'an},\ and\ \citenamefont
  {Font}}]{Miravet-Tenes:2023kte}%
  \BibitemOpen
  \bibfield  {author} {\bibinfo {author} {\bibfnamefont {M.}~\bibnamefont
  {Miravet-Ten\'es}}, \bibinfo {author} {\bibfnamefont {F.~L.}\ \bibnamefont
  {Castillo}}, \bibinfo {author} {\bibfnamefont {R.}~\bibnamefont {De~Pietri}},
  \bibinfo {author} {\bibfnamefont {P.}~\bibnamefont {Cerd\'a-Dur\'an}}, \ and\
  \bibinfo {author} {\bibfnamefont {J.~A.}\ \bibnamefont {Font}},\ }\href
  {\doibase 10.1103/PhysRevD.107.103053} {\bibfield  {journal} {\bibinfo
  {journal} {Phys. Rev. D}\ }\textbf {\bibinfo {volume} {107}},\ \bibinfo
  {pages} {103053} (\bibinfo {year} {2023})},\ \Eprint
  {http://arxiv.org/abs/2302.04553} {arXiv:2302.04553 [gr-qc]} \BibitemShut
  {NoStop}%
\bibitem [{\citenamefont {Raithel}\ \emph {et~al.}(2021)\citenamefont
  {Raithel}, \citenamefont {Paschalidis},\ and\ \citenamefont
  {\"Ozel}}]{Raithel:2021hye}%
  \BibitemOpen
  \bibfield  {author} {\bibinfo {author} {\bibfnamefont {C.}~\bibnamefont
  {Raithel}}, \bibinfo {author} {\bibfnamefont {V.}~\bibnamefont
  {Paschalidis}}, \ and\ \bibinfo {author} {\bibfnamefont {F.}~\bibnamefont
  {\"Ozel}},\ }\href {\doibase 10.1103/PhysRevD.104.063016} {\bibfield
  {journal} {\bibinfo  {journal} {Phys. Rev. D}\ }\textbf {\bibinfo {volume}
  {104}},\ \bibinfo {pages} {063016} (\bibinfo {year} {2021})},\ \Eprint
  {http://arxiv.org/abs/2104.07226} {arXiv:2104.07226 [astro-ph.HE]}
  \BibitemShut {NoStop}%
\bibitem [{\citenamefont {Chatziioannou}\ \emph {et~al.}(2017)\citenamefont
  {Chatziioannou}, \citenamefont {Clark}, \citenamefont {Bauswein},
  \citenamefont {Millhouse}, \citenamefont {Littenberg},\ and\ \citenamefont
  {Cornish}}]{Chatziioannou:2017ixj}%
  \BibitemOpen
  \bibfield  {author} {\bibinfo {author} {\bibfnamefont {K.}~\bibnamefont
  {Chatziioannou}}, \bibinfo {author} {\bibfnamefont {J.~A.}\ \bibnamefont
  {Clark}}, \bibinfo {author} {\bibfnamefont {A.}~\bibnamefont {Bauswein}},
  \bibinfo {author} {\bibfnamefont {M.}~\bibnamefont {Millhouse}}, \bibinfo
  {author} {\bibfnamefont {T.~B.}\ \bibnamefont {Littenberg}}, \ and\ \bibinfo
  {author} {\bibfnamefont {N.}~\bibnamefont {Cornish}},\ }\href {\doibase
  10.1103/PhysRevD.96.124035} {\bibfield  {journal} {\bibinfo  {journal} {Phys.
  Rev. D}\ }\textbf {\bibinfo {volume} {96}},\ \bibinfo {pages} {124035}
  (\bibinfo {year} {2017})},\ \Eprint {http://arxiv.org/abs/1711.00040}
  {arXiv:1711.00040 [gr-qc]} \BibitemShut {NoStop}%
\bibitem [{\citenamefont {Baiotti}\ and\ \citenamefont
  {Rezzolla}(2017)}]{Baiotti:2016qnr}%
  \BibitemOpen
  \bibfield  {author} {\bibinfo {author} {\bibfnamefont {L.}~\bibnamefont
  {Baiotti}}\ and\ \bibinfo {author} {\bibfnamefont {L.}~\bibnamefont
  {Rezzolla}},\ }\href {\doibase 10.1088/1361-6633/aa67bb} {\bibfield
  {journal} {\bibinfo  {journal} {Rept. Prog. Phys.}\ }\textbf {\bibinfo
  {volume} {80}},\ \bibinfo {pages} {096901} (\bibinfo {year} {2017})},\
  \Eprint {http://arxiv.org/abs/1607.03540} {arXiv:1607.03540 [gr-qc]}
  \BibitemShut {NoStop}%
\bibitem [{\citenamefont {Shibata}\ and\ \citenamefont
  {Hotokezaka}(2019)}]{Shibata:2019wef}%
  \BibitemOpen
  \bibfield  {author} {\bibinfo {author} {\bibfnamefont {M.}~\bibnamefont
  {Shibata}}\ and\ \bibinfo {author} {\bibfnamefont {K.}~\bibnamefont
  {Hotokezaka}},\ }\href {\doibase 10.1146/annurev-nucl-101918-023625}
  {\bibfield  {journal} {\bibinfo  {journal} {Ann. Rev. Nucl. Part. Sci.}\
  }\textbf {\bibinfo {volume} {69}},\ \bibinfo {pages} {41} (\bibinfo {year}
  {2019})},\ \Eprint {http://arxiv.org/abs/1908.02350} {arXiv:1908.02350
  [astro-ph.HE]} \BibitemShut {NoStop}%
\bibitem [{\citenamefont {Sarin}\ and\ \citenamefont
  {Lasky}(2021)}]{Sarin:2020gxb}%
  \BibitemOpen
  \bibfield  {author} {\bibinfo {author} {\bibfnamefont {N.}~\bibnamefont
  {Sarin}}\ and\ \bibinfo {author} {\bibfnamefont {P.~D.}\ \bibnamefont
  {Lasky}},\ }\href {\doibase 10.1007/s10714-021-02831-1} {\bibfield  {journal}
  {\bibinfo  {journal} {Gen. Rel. Grav.}\ }\textbf {\bibinfo {volume} {53}},\
  \bibinfo {pages} {59} (\bibinfo {year} {2021})},\ \Eprint
  {http://arxiv.org/abs/2012.08172} {arXiv:2012.08172 [astro-ph.HE]}
  \BibitemShut {NoStop}%
\bibitem [{\citenamefont {{Bauswein}}\ \emph {et~al.}(2020)\citenamefont
  {{Bauswein}}, \citenamefont {{Blacker}}, \citenamefont {{Vijayan}},
  \citenamefont {{Stergioulas}}, \citenamefont {{Chatziioannou}}, \citenamefont
  {{Clark}}, \citenamefont {{Bastian}}, \citenamefont {{Blaschke}},
  \citenamefont {{Cierniak}},\ and\ \citenamefont
  {{Fischer}}}]{2020PhRvL.125n1103B}%
  \BibitemOpen
  \bibfield  {author} {\bibinfo {author} {\bibfnamefont {A.}~\bibnamefont
  {{Bauswein}}}, \bibinfo {author} {\bibfnamefont {S.}~\bibnamefont
  {{Blacker}}}, \bibinfo {author} {\bibfnamefont {V.}~\bibnamefont
  {{Vijayan}}}, \bibinfo {author} {\bibfnamefont {N.}~\bibnamefont
  {{Stergioulas}}}, \bibinfo {author} {\bibfnamefont {K.}~\bibnamefont
  {{Chatziioannou}}}, \bibinfo {author} {\bibfnamefont {J.~A.}\ \bibnamefont
  {{Clark}}}, \bibinfo {author} {\bibfnamefont {N.-U.~F.}\ \bibnamefont
  {{Bastian}}}, \bibinfo {author} {\bibfnamefont {D.~B.}\ \bibnamefont
  {{Blaschke}}}, \bibinfo {author} {\bibfnamefont {M.}~\bibnamefont
  {{Cierniak}}}, \ and\ \bibinfo {author} {\bibfnamefont {T.}~\bibnamefont
  {{Fischer}}},\ }\href {\doibase 10.1103/PhysRevLett.125.141103} {\bibfield
  {journal} {\bibinfo  {journal} {\prl}\ }\textbf {\bibinfo {volume} {125}},\
  \bibinfo {eid} {141103} (\bibinfo {year} {2020})},\ \Eprint
  {http://arxiv.org/abs/2004.00846} {arXiv:2004.00846 [astro-ph.HE]}
  \BibitemShut {NoStop}%
\bibitem [{\citenamefont {Ruiz}\ \emph
  {et~al.}(2021{\natexlab{a}})\citenamefont {Ruiz}, \citenamefont {Shapiro},\
  and\ \citenamefont {Tsokaros}}]{Ruiz:2021gsv}%
  \BibitemOpen
  \bibfield  {author} {\bibinfo {author} {\bibfnamefont {M.}~\bibnamefont
  {Ruiz}}, \bibinfo {author} {\bibfnamefont {S.~L.}\ \bibnamefont {Shapiro}}, \
  and\ \bibinfo {author} {\bibfnamefont {A.}~\bibnamefont {Tsokaros}},\ }\href
  {\doibase 10.3389/fspas.2021.656907} {\bibfield  {journal} {\bibinfo
  {journal} {Front. Astron. Space Sci.}\ }\textbf {\bibinfo {volume} {8}},\
  \bibinfo {pages} {39} (\bibinfo {year} {2021}{\natexlab{a}})},\ \Eprint
  {http://arxiv.org/abs/2102.03366} {arXiv:2102.03366 [astro-ph.HE]}
  \BibitemShut {NoStop}%
\bibitem [{\citenamefont {{Janka}}\ \emph {et~al.}(1993)\citenamefont
  {{Janka}}, \citenamefont {{Zwerger}},\ and\ \citenamefont
  {{Moenchmeyer}}}]{1993A&A...268..360J}%
  \BibitemOpen
  \bibfield  {author} {\bibinfo {author} {\bibfnamefont {H.~T.}\ \bibnamefont
  {{Janka}}}, \bibinfo {author} {\bibfnamefont {T.}~\bibnamefont {{Zwerger}}},
  \ and\ \bibinfo {author} {\bibfnamefont {R.}~\bibnamefont {{Moenchmeyer}}},\
  }\href@noop {} {\bibfield  {journal} {\bibinfo  {journal} {Astron.
  Astrophys.}\ }\textbf {\bibinfo {volume} {268}},\ \bibinfo {pages} {360}
  (\bibinfo {year} {1993})}\BibitemShut {NoStop}%
\bibitem [{\citenamefont {{Dimmelmeier}}\ \emph {et~al.}(2002)\citenamefont
  {{Dimmelmeier}}, \citenamefont {{Font}},\ and\ \citenamefont
  {{M{\"u}ller}}}]{2002A&A...388..917D}%
  \BibitemOpen
  \bibfield  {author} {\bibinfo {author} {\bibfnamefont {H.}~\bibnamefont
  {{Dimmelmeier}}}, \bibinfo {author} {\bibfnamefont {J.~A.}\ \bibnamefont
  {{Font}}}, \ and\ \bibinfo {author} {\bibfnamefont {E.}~\bibnamefont
  {{M{\"u}ller}}},\ }\href {\doibase 10.1051/0004-6361:20020563} {\bibfield
  {journal} {\bibinfo  {journal} {Astron. Astrophys.}\ }\textbf {\bibinfo
  {volume} {388}},\ \bibinfo {pages} {917} (\bibinfo {year} {2002})},\ \Eprint
  {http://arxiv.org/abs/astro-ph/0204288} {arXiv:astro-ph/0204288 [astro-ph]}
  \BibitemShut {NoStop}%
\bibitem [{\citenamefont {Read}\ \emph {et~al.}(2009)\citenamefont {Read},
  \citenamefont {Lackey}, \citenamefont {Owen},\ and\ \citenamefont
  {Friedman}}]{Read:2008iy}%
  \BibitemOpen
  \bibfield  {author} {\bibinfo {author} {\bibfnamefont {J.~S.}\ \bibnamefont
  {Read}}, \bibinfo {author} {\bibfnamefont {B.~D.}\ \bibnamefont {Lackey}},
  \bibinfo {author} {\bibfnamefont {B.~J.}\ \bibnamefont {Owen}}, \ and\
  \bibinfo {author} {\bibfnamefont {J.~L.}\ \bibnamefont {Friedman}},\ }\href
  {\doibase 10.1103/PhysRevD.79.124032} {\bibfield  {journal} {\bibinfo
  {journal} {Phys. Rev.}\ }\textbf {\bibinfo {volume} {D79}},\ \bibinfo {pages}
  {124032} (\bibinfo {year} {2009})}\BibitemShut {NoStop}%
\bibitem [{\citenamefont {Constantinou}\ \emph {et~al.}(2015)\citenamefont
  {Constantinou}, \citenamefont {Muccioli}, \citenamefont {Prakash},\ and\
  \citenamefont {Lattimer}}]{Constantinou:2015mna}%
  \BibitemOpen
  \bibfield  {author} {\bibinfo {author} {\bibfnamefont {C.}~\bibnamefont
  {Constantinou}}, \bibinfo {author} {\bibfnamefont {B.}~\bibnamefont
  {Muccioli}}, \bibinfo {author} {\bibfnamefont {M.}~\bibnamefont {Prakash}}, \
  and\ \bibinfo {author} {\bibfnamefont {J.~M.}\ \bibnamefont {Lattimer}},\
  }\href {\doibase 10.1103/PhysRevC.92.025801} {\bibfield  {journal} {\bibinfo
  {journal} {Phys. Rev. C}\ }\textbf {\bibinfo {volume} {92}},\ \bibinfo
  {pages} {025801} (\bibinfo {year} {2015})},\ \Eprint
  {http://arxiv.org/abs/1504.03982} {arXiv:1504.03982 [astro-ph.SR]}
  \BibitemShut {NoStop}%
\bibitem [{\citenamefont {Lim}\ and\ \citenamefont {Holt}(2019)}]{Lim:2019ozm}%
  \BibitemOpen
  \bibfield  {author} {\bibinfo {author} {\bibfnamefont {Y.}~\bibnamefont
  {Lim}}\ and\ \bibinfo {author} {\bibfnamefont {J.~W.}\ \bibnamefont {Holt}},\
  }\href@noop {} {\  (\bibinfo {year} {2019})},\ \Eprint
  {http://arxiv.org/abs/1909.09089} {arXiv:1909.09089 [nucl-th]} \BibitemShut
  {NoStop}%
\bibitem [{\citenamefont {Bauswein}\ \emph {et~al.}(2010)\citenamefont
  {Bauswein}, \citenamefont {Janka},\ and\ \citenamefont
  {Oechslin}}]{Bauswein:2010dn}%
  \BibitemOpen
  \bibfield  {author} {\bibinfo {author} {\bibfnamefont {A.}~\bibnamefont
  {Bauswein}}, \bibinfo {author} {\bibfnamefont {H.~T.}\ \bibnamefont {Janka}},
  \ and\ \bibinfo {author} {\bibfnamefont {R.}~\bibnamefont {Oechslin}},\
  }\href {\doibase 10.1103/PhysRevD.82.084043} {\bibfield  {journal} {\bibinfo
  {journal} {Phys. Rev. D}\ }\textbf {\bibinfo {volume} {82}},\ \bibinfo
  {pages} {084043} (\bibinfo {year} {2010})},\ \Eprint
  {http://arxiv.org/abs/1006.3315} {arXiv:1006.3315 [astro-ph.SR]} \BibitemShut
  {NoStop}%
\bibitem [{\citenamefont {Figura}\ \emph {et~al.}(2021)\citenamefont {Figura},
  \citenamefont {Li}, \citenamefont {Lu}, \citenamefont {Burgio}, \citenamefont
  {Li},\ and\ \citenamefont {Schulze}}]{Figura:2021bcn}%
  \BibitemOpen
  \bibfield  {author} {\bibinfo {author} {\bibfnamefont {A.}~\bibnamefont
  {Figura}}, \bibinfo {author} {\bibfnamefont {F.}~\bibnamefont {Li}}, \bibinfo
  {author} {\bibfnamefont {J.-J.}\ \bibnamefont {Lu}}, \bibinfo {author}
  {\bibfnamefont {G.~F.}\ \bibnamefont {Burgio}}, \bibinfo {author}
  {\bibfnamefont {Z.-H.}\ \bibnamefont {Li}}, \ and\ \bibinfo {author}
  {\bibfnamefont {H.~J.}\ \bibnamefont {Schulze}},\ }\href {\doibase
  10.1103/PhysRevD.103.083012} {\bibfield  {journal} {\bibinfo  {journal}
  {Phys. Rev. D}\ }\textbf {\bibinfo {volume} {103}},\ \bibinfo {pages}
  {083012} (\bibinfo {year} {2021})},\ \Eprint
  {http://arxiv.org/abs/2103.02365} {arXiv:2103.02365 [gr-qc]} \BibitemShut
  {NoStop}%
\bibitem [{\citenamefont {Oechslin}\ \emph {et~al.}(2007)\citenamefont
  {Oechslin}, \citenamefont {Janka},\ and\ \citenamefont
  {Marek}}]{Oechslin:2006uk}%
  \BibitemOpen
  \bibfield  {author} {\bibinfo {author} {\bibfnamefont {R.}~\bibnamefont
  {Oechslin}}, \bibinfo {author} {\bibfnamefont {H.~T.}\ \bibnamefont {Janka}},
  \ and\ \bibinfo {author} {\bibfnamefont {A.}~\bibnamefont {Marek}},\ }\href
  {\doibase 10.1051/0004-6361:20066682} {\bibfield  {journal} {\bibinfo
  {journal} {Astron. Astrophys.}\ }\textbf {\bibinfo {volume} {467}},\ \bibinfo
  {pages} {395} (\bibinfo {year} {2007})},\ \Eprint
  {http://arxiv.org/abs/astro-ph/0611047} {arXiv:astro-ph/0611047} \BibitemShut
  {NoStop}%
\bibitem [{\citenamefont {Sekiguchi}\ \emph {et~al.}(2011)\citenamefont
  {Sekiguchi}, \citenamefont {Kiuchi}, \citenamefont {Kyutoku},\ and\
  \citenamefont {Shibata}}]{Sekiguchi:2011zd}%
  \BibitemOpen
  \bibfield  {author} {\bibinfo {author} {\bibfnamefont {Y.}~\bibnamefont
  {Sekiguchi}}, \bibinfo {author} {\bibfnamefont {K.}~\bibnamefont {Kiuchi}},
  \bibinfo {author} {\bibfnamefont {K.}~\bibnamefont {Kyutoku}}, \ and\
  \bibinfo {author} {\bibfnamefont {M.}~\bibnamefont {Shibata}},\ }\href
  {\doibase 10.1103/PhysRevLett.107.051102} {\bibfield  {journal} {\bibinfo
  {journal} {Phys. Rev. Lett.}\ }\textbf {\bibinfo {volume} {107}},\ \bibinfo
  {pages} {051102} (\bibinfo {year} {2011})},\ \Eprint
  {http://arxiv.org/abs/1105.2125} {arXiv:1105.2125 [gr-qc]} \BibitemShut
  {NoStop}%
\bibitem [{\citenamefont {Fields}\ \emph {et~al.}(2023)\citenamefont {Fields},
  \citenamefont {Prakash}, \citenamefont {Breschi}, \citenamefont {Radice},
  \citenamefont {Bernuzzi},\ and\ \citenamefont
  {da~Silva~Schneider}}]{Fields:2023bhs}%
  \BibitemOpen
  \bibfield  {author} {\bibinfo {author} {\bibfnamefont {J.}~\bibnamefont
  {Fields}}, \bibinfo {author} {\bibfnamefont {A.}~\bibnamefont {Prakash}},
  \bibinfo {author} {\bibfnamefont {M.}~\bibnamefont {Breschi}}, \bibinfo
  {author} {\bibfnamefont {D.}~\bibnamefont {Radice}}, \bibinfo {author}
  {\bibfnamefont {S.}~\bibnamefont {Bernuzzi}}, \ and\ \bibinfo {author}
  {\bibfnamefont {A.}~\bibnamefont {da~Silva~Schneider}},\ }\href {\doibase
  10.3847/2041-8213/ace5b2} {\bibfield  {journal} {\bibinfo  {journal}
  {Astrophys. J. Lett.}\ }\textbf {\bibinfo {volume} {952}},\ \bibinfo {pages}
  {L36} (\bibinfo {year} {2023})},\ \Eprint {http://arxiv.org/abs/2302.11359}
  {arXiv:2302.11359 [astro-ph.HE]} \BibitemShut {NoStop}%
\bibitem [{\citenamefont {Espino}\ \emph {et~al.}(2022)\citenamefont {Espino},
  \citenamefont {Bozzola},\ and\ \citenamefont {Paschalidis}}]{Espino:2022mtb}%
  \BibitemOpen
  \bibfield  {author} {\bibinfo {author} {\bibfnamefont {P.~L.}\ \bibnamefont
  {Espino}}, \bibinfo {author} {\bibfnamefont {G.}~\bibnamefont {Bozzola}}, \
  and\ \bibinfo {author} {\bibfnamefont {V.}~\bibnamefont {Paschalidis}},\
  }\href@noop {} {\  (\bibinfo {year} {2022})},\ \Eprint
  {http://arxiv.org/abs/2210.13481} {arXiv:2210.13481 [gr-qc]} \BibitemShut
  {NoStop}%
\bibitem [{\citenamefont {Werneck}\ \emph {et~al.}(2023)\citenamefont {Werneck}
  \emph {et~al.}}]{Werneck:2022exo}%
  \BibitemOpen
  \bibfield  {author} {\bibinfo {author} {\bibfnamefont {L.~R.}\ \bibnamefont
  {Werneck}} \emph {et~al.},\ }\href {\doibase 10.1103/PhysRevD.107.044037}
  {\bibfield  {journal} {\bibinfo  {journal} {Phys. Rev. D}\ }\textbf {\bibinfo
  {volume} {107}},\ \bibinfo {pages} {044037} (\bibinfo {year} {2023})},\
  \Eprint {http://arxiv.org/abs/2208.14487} {arXiv:2208.14487 [gr-qc]}
  \BibitemShut {NoStop}%
\bibitem [{\citenamefont {Zappa}\ \emph {et~al.}(2023)\citenamefont {Zappa},
  \citenamefont {Bernuzzi}, \citenamefont {Radice},\ and\ \citenamefont
  {Perego}}]{Zappa:2022rpd}%
  \BibitemOpen
  \bibfield  {author} {\bibinfo {author} {\bibfnamefont {F.}~\bibnamefont
  {Zappa}}, \bibinfo {author} {\bibfnamefont {S.}~\bibnamefont {Bernuzzi}},
  \bibinfo {author} {\bibfnamefont {D.}~\bibnamefont {Radice}}, \ and\ \bibinfo
  {author} {\bibfnamefont {A.}~\bibnamefont {Perego}},\ }\href {\doibase
  10.1093/mnras/stad107} {\bibfield  {journal} {\bibinfo  {journal} {Mon. Not.
  Roy. Astron. Soc.}\ }\textbf {\bibinfo {volume} {520}},\ \bibinfo {pages}
  {1481} (\bibinfo {year} {2023})},\ \Eprint {http://arxiv.org/abs/2210.11491}
  {arXiv:2210.11491 [astro-ph.HE]} \BibitemShut {NoStop}%
\bibitem [{\citenamefont {Blacker}\ \emph {et~al.}(2024)\citenamefont
  {Blacker}, \citenamefont {Kochankovski}, \citenamefont {Bauswein},
  \citenamefont {Ramos},\ and\ \citenamefont {Tolos}}]{Blacker:2023opp}%
  \BibitemOpen
  \bibfield  {author} {\bibinfo {author} {\bibfnamefont {S.}~\bibnamefont
  {Blacker}}, \bibinfo {author} {\bibfnamefont {H.}~\bibnamefont
  {Kochankovski}}, \bibinfo {author} {\bibfnamefont {A.}~\bibnamefont
  {Bauswein}}, \bibinfo {author} {\bibfnamefont {A.}~\bibnamefont {Ramos}}, \
  and\ \bibinfo {author} {\bibfnamefont {L.}~\bibnamefont {Tolos}},\ }\href
  {\doibase 10.1103/PhysRevD.109.043015} {\bibfield  {journal} {\bibinfo
  {journal} {Phys. Rev. D}\ }\textbf {\bibinfo {volume} {109}},\ \bibinfo
  {pages} {043015} (\bibinfo {year} {2024})},\ \Eprint
  {http://arxiv.org/abs/2307.03710} {arXiv:2307.03710 [astro-ph.HE]}
  \BibitemShut {NoStop}%
\bibitem [{\citenamefont {Bamber}\ \emph {et~al.}(2024)\citenamefont {Bamber},
  \citenamefont {Tsokaros}, \citenamefont {Ruiz},\ and\ \citenamefont
  {Shapiro}}]{Bamber:2024kfb}%
  \BibitemOpen
  \bibfield  {author} {\bibinfo {author} {\bibfnamefont {J.}~\bibnamefont
  {Bamber}}, \bibinfo {author} {\bibfnamefont {A.}~\bibnamefont {Tsokaros}},
  \bibinfo {author} {\bibfnamefont {M.}~\bibnamefont {Ruiz}}, \ and\ \bibinfo
  {author} {\bibfnamefont {S.~L.}\ \bibnamefont {Shapiro}},\ }\href {\doibase
  10.1103/PhysRevD.110.024046} {\bibfield  {journal} {\bibinfo  {journal}
  {Phys. Rev. D}\ }\textbf {\bibinfo {volume} {110}},\ \bibinfo {pages}
  {024046} (\bibinfo {year} {2024})},\ \Eprint
  {http://arxiv.org/abs/2405.03705} {arXiv:2405.03705 [astro-ph.HE]}
  \BibitemShut {NoStop}%
\bibitem [{\citenamefont {Tsokaros}\ \emph {et~al.}(2025)\citenamefont
  {Tsokaros}, \citenamefont {Bamber}, \citenamefont {Ruiz},\ and\ \citenamefont
  {Shapiro}}]{Tsokaros:2024wgb}%
  \BibitemOpen
  \bibfield  {author} {\bibinfo {author} {\bibfnamefont {A.}~\bibnamefont
  {Tsokaros}}, \bibinfo {author} {\bibfnamefont {J.}~\bibnamefont {Bamber}},
  \bibinfo {author} {\bibfnamefont {M.}~\bibnamefont {Ruiz}}, \ and\ \bibinfo
  {author} {\bibfnamefont {S.~L.}\ \bibnamefont {Shapiro}},\ }\href {\doibase
  10.1103/PhysRevLett.134.121401} {\bibfield  {journal} {\bibinfo  {journal}
  {Phys. Rev. Lett.}\ }\textbf {\bibinfo {volume} {134}},\ \bibinfo {pages}
  {121401} (\bibinfo {year} {2025})},\ \Eprint
  {http://arxiv.org/abs/2411.00939} {arXiv:2411.00939 [gr-qc]} \BibitemShut
  {NoStop}%
\bibitem [{\citenamefont {stellarcollapse}(2022)}]{stellarcollapse}%
  \BibitemOpen
  \bibfield  {author} {\bibinfo {author} {\bibnamefont {stellarcollapse}},\
  }\href@noop {} {} (\bibinfo {year} {Accessed in 2022}),\ \bibinfo {note}
  {\url{https://stellarcollapse.org/}}\BibitemShut {NoStop}%
\bibitem [{\citenamefont {Raithel}\ \emph {et~al.}(2019)\citenamefont
  {Raithel}, \citenamefont {Ozel},\ and\ \citenamefont
  {Psaltis}}]{Raithel:2019gws}%
  \BibitemOpen
  \bibfield  {author} {\bibinfo {author} {\bibfnamefont {C.~A.}\ \bibnamefont
  {Raithel}}, \bibinfo {author} {\bibfnamefont {F.}~\bibnamefont {Ozel}}, \
  and\ \bibinfo {author} {\bibfnamefont {D.}~\bibnamefont {Psaltis}},\ }\href
  {\doibase 10.3847/1538-4357/ab08ea} {\bibfield  {journal} {\bibinfo
  {journal} {Astrophys. J.}\ }\textbf {\bibinfo {volume} {875}},\ \bibinfo
  {pages} {12} (\bibinfo {year} {2019})},\ \Eprint
  {http://arxiv.org/abs/1902.10735} {arXiv:1902.10735 [astro-ph.HE]}
  \BibitemShut {NoStop}%
\bibitem [{\citenamefont {Mroczek}\ \emph {et~al.}(2024)\citenamefont
  {Mroczek}, \citenamefont {Yao}, \citenamefont {Zine}, \citenamefont
  {Noronha-Hostler}, \citenamefont {Dexheimer}, \citenamefont {Haber},\ and\
  \citenamefont {Most}}]{Mroczek:2024sfp}%
  \BibitemOpen
  \bibfield  {author} {\bibinfo {author} {\bibfnamefont {D.}~\bibnamefont
  {Mroczek}}, \bibinfo {author} {\bibfnamefont {N.}~\bibnamefont {Yao}},
  \bibinfo {author} {\bibfnamefont {K.}~\bibnamefont {Zine}}, \bibinfo {author}
  {\bibfnamefont {J.}~\bibnamefont {Noronha-Hostler}}, \bibinfo {author}
  {\bibfnamefont {V.}~\bibnamefont {Dexheimer}}, \bibinfo {author}
  {\bibfnamefont {A.}~\bibnamefont {Haber}}, \ and\ \bibinfo {author}
  {\bibfnamefont {E.~R.}\ \bibnamefont {Most}},\ }\href@noop {} {\  (\bibinfo
  {year} {2024})},\ \Eprint {http://arxiv.org/abs/2404.01658} {arXiv:2404.01658
  [astro-ph.HE]} \BibitemShut {NoStop}%
\bibitem [{\citenamefont {Guerra}\ \emph {et~al.}(2024)\citenamefont {Guerra}
  \emph {et~al.}}]{Guerra:2025}%
  \BibitemOpen
  \bibfield  {author} {\bibinfo {author} {\bibfnamefont {D.}~\bibnamefont
  {Guerra}} \emph {et~al.},\ }\href@noop {} {\bibfield  {journal} {\bibinfo
  {journal} {in preparation}\ } (\bibinfo {year} {2024})}\BibitemShut {NoStop}%
\bibitem [{\citenamefont {Bustillo}\ \emph {et~al.}(2022)\citenamefont
  {Bustillo}, \citenamefont {Wong}, \citenamefont {Sanchis-Gual}, \citenamefont
  {Leong}, \citenamefont {Torres-Forne}, \citenamefont {Chandra}, \citenamefont
  {Font}, \citenamefont {Herdeiro}, \citenamefont {Radu},\ and\ \citenamefont
  {Li}}]{Psi4PE}%
  \BibitemOpen
  \bibfield  {author} {\bibinfo {author} {\bibfnamefont {J.~C.}\ \bibnamefont
  {Bustillo}}, \bibinfo {author} {\bibfnamefont {I.~C.~F.}\ \bibnamefont
  {Wong}}, \bibinfo {author} {\bibfnamefont {N.}~\bibnamefont {Sanchis-Gual}},
  \bibinfo {author} {\bibfnamefont {S.~H.~W.}\ \bibnamefont {Leong}}, \bibinfo
  {author} {\bibfnamefont {A.}~\bibnamefont {Torres-Forne}}, \bibinfo {author}
  {\bibfnamefont {K.}~\bibnamefont {Chandra}}, \bibinfo {author} {\bibfnamefont
  {J.~A.}\ \bibnamefont {Font}}, \bibinfo {author} {\bibfnamefont
  {C.}~\bibnamefont {Herdeiro}}, \bibinfo {author} {\bibfnamefont
  {E.}~\bibnamefont {Radu}}, \ and\ \bibinfo {author} {\bibfnamefont
  {T.~G.~F.}\ \bibnamefont {Li}},\ }\href@noop {} {\enquote {\bibinfo {title}
  {Gravitational-wave parameter inference with the newman-penrose scalar},}\ }
  (\bibinfo {year} {2022}),\ \Eprint {http://arxiv.org/abs/arXiv:2205.15029}
  {arXiv:2205.15029} \BibitemShut {NoStop}%
\bibitem [{\citenamefont {Reisswig}\ and\ \citenamefont
  {Pollney}(2010)}]{Pollney_Reissweig}%
  \BibitemOpen
  \bibfield  {author} {\bibinfo {author} {\bibfnamefont {C.}~\bibnamefont
  {Reisswig}}\ and\ \bibinfo {author} {\bibfnamefont {D.}~\bibnamefont
  {Pollney}},\ }\href {\doibase 10.1088/0264-9381/28/19/195015} {\  (\bibinfo
  {year} {2010}),\ 10.1088/0264-9381/28/19/195015},\ \Eprint
  {http://arxiv.org/abs/arXiv:1006.1632} {arXiv:1006.1632} \BibitemShut
  {NoStop}%
\bibitem [{\citenamefont {Miravet-Ten\'es}\ \emph {et~al.}(2024)\citenamefont
  {Miravet-Ten\'es}, \citenamefont {Guerra}, \citenamefont {Ruiz},
  \citenamefont {Cerd\'a-Dur\'an},\ and\ \citenamefont
  {Font}}]{Miravet-Tenes:2023}%
  \BibitemOpen
  \bibfield  {author} {\bibinfo {author} {\bibfnamefont {M.}~\bibnamefont
  {Miravet-Ten\'es}}, \bibinfo {author} {\bibfnamefont {D.}~\bibnamefont
  {Guerra}}, \bibinfo {author} {\bibfnamefont {M.}~\bibnamefont {Ruiz}},
  \bibinfo {author} {\bibfnamefont {P.}~\bibnamefont {Cerd\'a-Dur\'an}}, \ and\
  \bibinfo {author} {\bibfnamefont {J.~A.}\ \bibnamefont {Font}},\ }\href@noop
  {} {\  (\bibinfo {year} {2024})},\ \Eprint {http://arxiv.org/abs/2401.02493}
  {arXiv:2401.02493 [gr-qc]} \BibitemShut {NoStop}%
\bibitem [{\citenamefont {{Cornish}}\ and\ \citenamefont
  {{Littenberg}}(2015)}]{Cornish:2015}%
  \BibitemOpen
  \bibfield  {author} {\bibinfo {author} {\bibfnamefont {N.~J.}\ \bibnamefont
  {{Cornish}}}\ and\ \bibinfo {author} {\bibfnamefont {T.~B.}\ \bibnamefont
  {{Littenberg}}},\ }\href {\doibase 10.1088/0264-9381/32/13/135012} {\bibfield
   {journal} {\bibinfo  {journal} {Classical and Quantum Gravity}\ }\textbf
  {\bibinfo {volume} {32}},\ \bibinfo {eid} {135012} (\bibinfo {year}
  {2015})},\ \Eprint {http://arxiv.org/abs/1410.3835} {arXiv:1410.3835 [gr-qc]}
  \BibitemShut {NoStop}%
\bibitem [{\citenamefont {Schneider}\ \emph {et~al.}(2017)\citenamefont
  {Schneider}, \citenamefont {Roberts},\ and\ \citenamefont
  {Ott}}]{Schneider:2017tfi}%
  \BibitemOpen
  \bibfield  {author} {\bibinfo {author} {\bibfnamefont {A.~S.}\ \bibnamefont
  {Schneider}}, \bibinfo {author} {\bibfnamefont {L.~F.}\ \bibnamefont
  {Roberts}}, \ and\ \bibinfo {author} {\bibfnamefont {C.~D.}\ \bibnamefont
  {Ott}},\ }\href {\doibase 10.1103/PhysRevC.96.065802} {\bibfield  {journal}
  {\bibinfo  {journal} {Phys. Rev. C}\ }\textbf {\bibinfo {volume} {96}},\
  \bibinfo {pages} {065802} (\bibinfo {year} {2017})},\ \Eprint
  {http://arxiv.org/abs/1707.01527} {arXiv:1707.01527 [astro-ph.HE]}
  \BibitemShut {NoStop}%
\bibitem [{\citenamefont {{Pilgrim}}(2021)}]{2021JOSS....6.3859P}%
  \BibitemOpen
  \bibfield  {author} {\bibinfo {author} {\bibfnamefont {C.}~\bibnamefont
  {{Pilgrim}}},\ }\href {\doibase 10.21105/joss.03859} {\bibfield  {journal}
  {\bibinfo  {journal} {The Journal of Open Source Software}\ }\textbf
  {\bibinfo {volume} {6}},\ \bibinfo {eid} {3859} (\bibinfo {year}
  {2021})}\BibitemShut {NoStop}%
\bibitem [{\citenamefont {Gourgoulhon}\ \emph {et~al.}(2001)\citenamefont
  {Gourgoulhon}, \citenamefont {Grandclement}, \citenamefont {Taniguchi},
  \citenamefont {Marck},\ and\ \citenamefont {Bonazzola}}]{Gourgoulhon:2000nn}%
  \BibitemOpen
  \bibfield  {author} {\bibinfo {author} {\bibfnamefont {E.}~\bibnamefont
  {Gourgoulhon}}, \bibinfo {author} {\bibfnamefont {P.}~\bibnamefont
  {Grandclement}}, \bibinfo {author} {\bibfnamefont {K.}~\bibnamefont
  {Taniguchi}}, \bibinfo {author} {\bibfnamefont {J.-A.}\ \bibnamefont
  {Marck}}, \ and\ \bibinfo {author} {\bibfnamefont {S.}~\bibnamefont
  {Bonazzola}},\ }\href {\doibase 10.1103/PhysRevD.63.064029} {\bibfield
  {journal} {\bibinfo  {journal} {Phys. Rev. D}\ }\textbf {\bibinfo {volume}
  {63}},\ \bibinfo {pages} {064029} (\bibinfo {year} {2001})},\ \Eprint
  {http://arxiv.org/abs/gr-qc/0007028} {arXiv:gr-qc/0007028} \BibitemShut
  {NoStop}%
\bibitem [{\citenamefont {{Taniguchi}}\ and\ \citenamefont
  {{Gourgoulhon}}(2002)}]{tg02}%
  \BibitemOpen
  \bibfield  {author} {\bibinfo {author} {\bibfnamefont {K.}~\bibnamefont
  {{Taniguchi}}}\ and\ \bibinfo {author} {\bibfnamefont {E.}~\bibnamefont
  {{Gourgoulhon}}},\ }\href {\doibase 10.1103/PhysRevD.66.104019} {\bibfield
  {journal} {\bibinfo  {journal} {Phys. Rev. D}\ }\textbf {\bibinfo {volume}
  {66}},\ \bibinfo {eid} {104019} (\bibinfo {year} {2002})}\BibitemShut
  {NoStop}%
\bibitem [{\citenamefont {Etienne}\ \emph {et~al.}(2015)\citenamefont
  {Etienne}, \citenamefont {Paschalidis}, \citenamefont {Haas}, \citenamefont
  {M\"osta},\ and\ \citenamefont {Shapiro}}]{Etienne:2015cea}%
  \BibitemOpen
  \bibfield  {author} {\bibinfo {author} {\bibfnamefont {Z.~B.}\ \bibnamefont
  {Etienne}}, \bibinfo {author} {\bibfnamefont {V.}~\bibnamefont
  {Paschalidis}}, \bibinfo {author} {\bibfnamefont {R.}~\bibnamefont {Haas}},
  \bibinfo {author} {\bibfnamefont {P.}~\bibnamefont {M\"osta}}, \ and\
  \bibinfo {author} {\bibfnamefont {S.~L.}\ \bibnamefont {Shapiro}},\ }\href
  {\doibase 10.1088/0264-9381/32/17/175009} {\bibfield  {journal} {\bibinfo
  {journal} {Class. Quant. Grav.}\ }\textbf {\bibinfo {volume} {32}},\ \bibinfo
  {pages} {175009} (\bibinfo {year} {2015})},\ \Eprint
  {http://arxiv.org/abs/1501.07276} {arXiv:1501.07276 [astro-ph.HE]}
  \BibitemShut {NoStop}%
\bibitem [{\citenamefont {Loffler}\ \emph {et~al.}(2012)\citenamefont {Loffler}
  \emph {et~al.}}]{Loffler:2011ay}%
  \BibitemOpen
  \bibfield  {author} {\bibinfo {author} {\bibfnamefont {F.}~\bibnamefont
  {Loffler}} \emph {et~al.},\ }\href {\doibase 10.1088/0264-9381/29/11/115001}
  {\bibfield  {journal} {\bibinfo  {journal} {Class. Quant. Grav.}\ }\textbf
  {\bibinfo {volume} {29}},\ \bibinfo {pages} {115001} (\bibinfo {year}
  {2012})},\ \Eprint {http://arxiv.org/abs/1111.3344} {arXiv:1111.3344 [gr-qc]}
  \BibitemShut {NoStop}%
\bibitem [{\citenamefont {{Baumgarte}}\ and\ \citenamefont
  {{Shapiro}}(1998)}]{Baumgarte:1998}%
  \BibitemOpen
  \bibfield  {author} {\bibinfo {author} {\bibfnamefont {T.~W.}\ \bibnamefont
  {{Baumgarte}}}\ and\ \bibinfo {author} {\bibfnamefont {S.~L.}\ \bibnamefont
  {{Shapiro}}},\ }\href {\doibase 10.1103/PhysRevD.59.024007} {\bibfield
  {journal} {\bibinfo  {journal} {\prd}\ }\textbf {\bibinfo {volume} {59}},\
  \bibinfo {eid} {024007} (\bibinfo {year} {1998})},\ \Eprint
  {http://arxiv.org/abs/gr-qc/9810065} {arXiv:gr-qc/9810065 [gr-qc]}
  \BibitemShut {NoStop}%
\bibitem [{\citenamefont {Shibata}\ and\ \citenamefont
  {Nakamura}(1995)}]{Shibata:1995}%
  \BibitemOpen
  \bibfield  {author} {\bibinfo {author} {\bibfnamefont {M.}~\bibnamefont
  {Shibata}}\ and\ \bibinfo {author} {\bibfnamefont {T.}~\bibnamefont
  {Nakamura}},\ }\href {\doibase 10.1103/PhysRevD.52.5428} {\bibfield
  {journal} {\bibinfo  {journal} {Phys. Rev. D}\ }\textbf {\bibinfo {volume}
  {52}},\ \bibinfo {pages} {5428} (\bibinfo {year} {1995})}\BibitemShut
  {NoStop}%
\bibitem [{\citenamefont {{Banyuls}}\ \emph {et~al.}(1997)\citenamefont
  {{Banyuls}}, \citenamefont {{Font}}, \citenamefont {{Ib{\'a}{\~n}ez}},
  \citenamefont {{Mart{\'\i}}},\ and\ \citenamefont {{Miralles}}}]{Valencia}%
  \BibitemOpen
  \bibfield  {author} {\bibinfo {author} {\bibfnamefont {F.}~\bibnamefont
  {{Banyuls}}}, \bibinfo {author} {\bibfnamefont {J.~A.}\ \bibnamefont
  {{Font}}}, \bibinfo {author} {\bibfnamefont {J.~M.}\ \bibnamefont
  {{Ib{\'a}{\~n}ez}}}, \bibinfo {author} {\bibfnamefont {J.~M.}\ \bibnamefont
  {{Mart{\'\i}}}}, \ and\ \bibinfo {author} {\bibfnamefont {J.~A.}\
  \bibnamefont {{Miralles}}},\ }\href {\doibase 10.1086/303604} {\bibfield
  {journal} {\bibinfo  {journal} {\apj}\ }\textbf {\bibinfo {volume} {476}},\
  \bibinfo {pages} {221} (\bibinfo {year} {1997})}\BibitemShut {NoStop}%
\bibitem [{\citenamefont {Reitze}\ \emph {et~al.}(2019)\citenamefont {Reitze},
  \citenamefont {Adhikari}, \citenamefont {Ballmer}, \citenamefont {Barish},
  \citenamefont {Barsotti}, \citenamefont {Billingsley}, \citenamefont {Brown},
  \citenamefont {Chen}, \citenamefont {Coyne}, \citenamefont {Eisenstein},
  \citenamefont {Evans}, \citenamefont {Fritschel}, \citenamefont {Hall},
  \citenamefont {Lazzarini}, \citenamefont {Lovelace}, \citenamefont {Read},
  \citenamefont {Sathyaprakash}, \citenamefont {Shoemaker}, \citenamefont
  {Smith}, \citenamefont {Torrie}, \citenamefont {Vitale}, \citenamefont
  {Weiss}, \citenamefont {Wipf},\ and\ \citenamefont {Zucker}}]{CE}%
  \BibitemOpen
  \bibfield  {author} {\bibinfo {author} {\bibfnamefont {D.}~\bibnamefont
  {Reitze}}, \bibinfo {author} {\bibfnamefont {R.~X.}\ \bibnamefont
  {Adhikari}}, \bibinfo {author} {\bibfnamefont {S.}~\bibnamefont {Ballmer}},
  \bibinfo {author} {\bibfnamefont {B.}~\bibnamefont {Barish}}, \bibinfo
  {author} {\bibfnamefont {L.}~\bibnamefont {Barsotti}}, \bibinfo {author}
  {\bibfnamefont {G.}~\bibnamefont {Billingsley}}, \bibinfo {author}
  {\bibfnamefont {D.~A.}\ \bibnamefont {Brown}}, \bibinfo {author}
  {\bibfnamefont {Y.}~\bibnamefont {Chen}}, \bibinfo {author} {\bibfnamefont
  {D.}~\bibnamefont {Coyne}}, \bibinfo {author} {\bibfnamefont
  {R.}~\bibnamefont {Eisenstein}}, \bibinfo {author} {\bibfnamefont
  {M.}~\bibnamefont {Evans}}, \bibinfo {author} {\bibfnamefont
  {P.}~\bibnamefont {Fritschel}}, \bibinfo {author} {\bibfnamefont {E.~D.}\
  \bibnamefont {Hall}}, \bibinfo {author} {\bibfnamefont {A.}~\bibnamefont
  {Lazzarini}}, \bibinfo {author} {\bibfnamefont {G.}~\bibnamefont {Lovelace}},
  \bibinfo {author} {\bibfnamefont {J.}~\bibnamefont {Read}}, \bibinfo {author}
  {\bibfnamefont {B.~S.}\ \bibnamefont {Sathyaprakash}}, \bibinfo {author}
  {\bibfnamefont {D.}~\bibnamefont {Shoemaker}}, \bibinfo {author}
  {\bibfnamefont {J.}~\bibnamefont {Smith}}, \bibinfo {author} {\bibfnamefont
  {C.}~\bibnamefont {Torrie}}, \bibinfo {author} {\bibfnamefont
  {S.}~\bibnamefont {Vitale}}, \bibinfo {author} {\bibfnamefont
  {R.}~\bibnamefont {Weiss}}, \bibinfo {author} {\bibfnamefont
  {C.}~\bibnamefont {Wipf}}, \ and\ \bibinfo {author} {\bibfnamefont
  {M.}~\bibnamefont {Zucker}},\ }\href@noop {} {\  (\bibinfo {year} {2019})},\
  \Eprint {http://arxiv.org/abs/arXiv:1907.04833} {arXiv:1907.04833}
  \BibitemShut {NoStop}%
\bibitem [{\citenamefont {Abbott}\ \emph
  {et~al.}(2017{\natexlab{f}})\citenamefont {Abbott} \emph {et~al.}}]{CE2}%
  \BibitemOpen
  \bibfield  {author} {\bibinfo {author} {\bibfnamefont {B.~P.}\ \bibnamefont
  {Abbott}} \emph {et~al.},\ }\href {\doibase 10.1088/1361-6382/aa51f4}
  {\bibfield  {journal} {\bibinfo  {journal} {Classical and Quantum Gravity}\
  }\textbf {\bibinfo {volume} {34}},\ \bibinfo {pages} {044001} (\bibinfo
  {year} {2017}{\natexlab{f}})}\BibitemShut {NoStop}%
\bibitem [{\citenamefont {Collaboration}(2018)}]{LIGOsens}%
  \BibitemOpen
  \bibfield  {author} {\bibinfo {author} {\bibfnamefont {L.~S.}\ \bibnamefont
  {Collaboration}},\ }\href {\doibase https://doi.org/10.7935/GT1W-FZ16}
  {\bibfield  {journal} {\bibinfo  {journal} {free software (GPL)}\ } (\bibinfo
  {year} {2018}),\ https://doi.org/10.7935/GT1W-FZ16}\BibitemShut {NoStop}%
\bibitem [{\citenamefont {Acernese}(2015)}]{VIRGO:2014yos}%
  \BibitemOpen
  \bibfield  {author} {\bibinfo {author} {\bibfnamefont {F.~e.~a.}\
  \bibnamefont {Acernese}} (\bibinfo {collaboration} {VIRGO}),\ }\href
  {\doibase 10.1088/0264-9381/32/2/024001} {\bibfield  {journal} {\bibinfo
  {journal} {Class. Quant. Grav.}\ }\textbf {\bibinfo {volume} {32}},\ \bibinfo
  {pages} {024001} (\bibinfo {year} {2015})},\ \Eprint
  {http://arxiv.org/abs/1408.3978} {arXiv:1408.3978 [gr-qc]} \BibitemShut
  {NoStop}%
\bibitem [{\citenamefont {{Hild}}(2011)}]{2011CQGra..28i4013H}%
  \BibitemOpen
  \bibfield  {author} {\bibinfo {author} {\bibfnamefont {S.~e.~a.}\
  \bibnamefont {{Hild}}} (\bibinfo {collaboration} {ET}),\ }\href {\doibase
  10.1088/0264-9381/28/9/094013} {\bibfield  {journal} {\bibinfo  {journal}
  {Classical and Quantum Gravity}\ }\textbf {\bibinfo {volume} {28}},\ \bibinfo
  {eid} {094013} (\bibinfo {year} {2011})},\ \Eprint
  {http://arxiv.org/abs/1012.0908} {arXiv:1012.0908 [gr-qc]} \BibitemShut
  {NoStop}%
\bibitem [{\citenamefont {{Abbott}}\ \emph {et~al.}(2017)\citenamefont
  {{Abbott}}, \citenamefont {{(LIGO Scientific Collaboration}},\ and\
  \citenamefont {{Harms}}}]{2017CQGra..34d4001A}%
  \BibitemOpen
  \bibfield  {author} {\bibinfo {author} {\bibfnamefont {B.~P. e.~a.}\
  \bibnamefont {{Abbott}}}, \bibinfo {author} {\bibnamefont {{(LIGO Scientific
  Collaboration}}}, \ and\ \bibinfo {author} {\bibfnamefont {J.}~\bibnamefont
  {{Harms}}},\ }\href {\doibase 10.1088/1361-6382/aa51f4} {\bibfield  {journal}
  {\bibinfo  {journal} {Classical and Quantum Gravity}\ }\textbf {\bibinfo
  {volume} {34}},\ \bibinfo {eid} {044001} (\bibinfo {year} {2017})},\ \Eprint
  {http://arxiv.org/abs/1607.08697} {arXiv:1607.08697 [astro-ph.IM]}
  \BibitemShut {NoStop}%
\bibitem [{\citenamefont {Finn}(1992)}]{Finn1992}%
  \BibitemOpen
  \bibfield  {author} {\bibinfo {author} {\bibfnamefont {L.~S.}\ \bibnamefont
  {Finn}},\ }\href {\doibase 10.1103/physrevd.46.5236} {\bibfield  {journal}
  {\bibinfo  {journal} {Physical Review D}\ }\textbf {\bibinfo {volume} {46}},\
  \bibinfo {pages} {5236} (\bibinfo {year} {1992})}\BibitemShut {NoStop}%
\bibitem [{\citenamefont {Romano}\ and\ \citenamefont
  {Cornish}(2017)}]{Romano2017}%
  \BibitemOpen
  \bibfield  {author} {\bibinfo {author} {\bibfnamefont {J.~D.}\ \bibnamefont
  {Romano}}\ and\ \bibinfo {author} {\bibfnamefont {N.~J.}\ \bibnamefont
  {Cornish}},\ }\href {\doibase 10.1007/s41114-017-0004-1} {\bibfield
  {journal} {\bibinfo  {journal} {Living Reviews in Relativity}\ }\textbf
  {\bibinfo {volume} {20}} (\bibinfo {year} {2017}),\
  10.1007/s41114-017-0004-1}\BibitemShut {NoStop}%
\bibitem [{\citenamefont {Cutler}\ and\ \citenamefont
  {Flanagan}(1994)}]{Cutler1994}%
  \BibitemOpen
  \bibfield  {author} {\bibinfo {author} {\bibfnamefont {C.}~\bibnamefont
  {Cutler}}\ and\ \bibinfo {author} {\bibfnamefont {{\'{E}}.~E.}\ \bibnamefont
  {Flanagan}},\ }\href {\doibase 10.1103/physrevd.49.2658} {\bibfield
  {journal} {\bibinfo  {journal} {Physical Review D}\ }\textbf {\bibinfo
  {volume} {49}},\ \bibinfo {pages} {2658} (\bibinfo {year}
  {1994})}\BibitemShut {NoStop}%
\bibitem [{\citenamefont {Abbott}\ \emph {et~al.}(2017)\citenamefont {Abbott}
  \emph {et~al.}}]{ExploringSens}%
  \BibitemOpen
  \bibfield  {author} {\bibinfo {author} {\bibfnamefont {B.~P.}\ \bibnamefont
  {Abbott}} \emph {et~al.},\ }\href {\doibase 10.1088/1361-6382/aa51f4}
  {\bibfield  {journal} {\bibinfo  {journal} {Classical and Quantum Gravity}\
  }\textbf {\bibinfo {volume} {34}},\ \bibinfo {pages} {044001} (\bibinfo
  {year} {2017})}\BibitemShut {NoStop}%
\bibitem [{\citenamefont {Smith}\ \emph {et~al.}(2021)\citenamefont {Smith}
  \emph {et~al.}}]{Smith2021_BNS}%
  \BibitemOpen
  \bibfield  {author} {\bibinfo {author} {\bibfnamefont {R.}~\bibnamefont
  {Smith}} \emph {et~al.},\ }\href {\doibase 10.1103/physrevlett.127.081102}
  {\bibfield  {journal} {\bibinfo  {journal} {Physical Review Letters}\
  }\textbf {\bibinfo {volume} {127}} (\bibinfo {year} {2021}),\
  10.1103/physrevlett.127.081102}\BibitemShut {NoStop}%
\bibitem [{\citenamefont {Branchesi}\ \emph {et~al.}(2023)\citenamefont
  {Branchesi} \emph {et~al.}}]{Branchesi:2023mws}%
  \BibitemOpen
  \bibfield  {author} {\bibinfo {author} {\bibfnamefont {M.}~\bibnamefont
  {Branchesi}} \emph {et~al.},\ }\href {\doibase 10.1088/1475-7516/2023/07/068}
  {\bibfield  {journal} {\bibinfo  {journal} {JCAP}\ }\textbf {\bibinfo
  {volume} {07}},\ \bibinfo {pages} {068} (\bibinfo {year} {2023})},\ \Eprint
  {http://arxiv.org/abs/2303.15923} {arXiv:2303.15923 [gr-qc]} \BibitemShut
  {NoStop}%
\bibitem [{\citenamefont {Ashton}\ \emph {et~al.}(2019)\citenamefont {Ashton}
  \emph {et~al.}}]{Ashton:2018jfp}%
  \BibitemOpen
  \bibfield  {author} {\bibinfo {author} {\bibfnamefont {G.}~\bibnamefont
  {Ashton}} \emph {et~al.},\ }\href {\doibase 10.3847/1538-4365/ab06fc}
  {\bibfield  {journal} {\bibinfo  {journal} {Astrophys. J. Suppl.}\ }\textbf
  {\bibinfo {volume} {241}},\ \bibinfo {pages} {27} (\bibinfo {year} {2019})},\
  \Eprint {http://arxiv.org/abs/1811.02042} {arXiv:1811.02042 [astro-ph.IM]}
  \BibitemShut {NoStop}%
\bibitem [{\citenamefont {Smith}\ \emph {et~al.}(2020)\citenamefont {Smith},
  \citenamefont {Ashton}, \citenamefont {Vajpeyi},\ and\ \citenamefont
  {Talbot}}]{pbilby}%
  \BibitemOpen
  \bibfield  {author} {\bibinfo {author} {\bibfnamefont {R.~J.~E.}\
  \bibnamefont {Smith}}, \bibinfo {author} {\bibfnamefont {G.}~\bibnamefont
  {Ashton}}, \bibinfo {author} {\bibfnamefont {A.}~\bibnamefont {Vajpeyi}}, \
  and\ \bibinfo {author} {\bibfnamefont {C.}~\bibnamefont {Talbot}},\ }\href
  {\doibase 10.1093/mnras/staa2483} {\bibfield  {journal} {\bibinfo  {journal}
  {Monthly Notices of the Royal Astronomical Society}\ }\textbf {\bibinfo
  {volume} {498}},\ \bibinfo {pages} {4492} (\bibinfo {year}
  {2020})}\BibitemShut {NoStop}%
\bibitem [{\citenamefont {Speagle}(2020)}]{Dynesty}%
  \BibitemOpen
  \bibfield  {author} {\bibinfo {author} {\bibfnamefont {J.~S.}\ \bibnamefont
  {Speagle}},\ }\href {\doibase 10.1093/mnras/staa278} {\bibfield  {journal}
  {\bibinfo  {journal} {Monthly Notices of the Royal Astronomical Society}\
  }\textbf {\bibinfo {volume} {493}},\ \bibinfo {pages} {3132} (\bibinfo {year}
  {2020})}\BibitemShut {NoStop}%
\bibitem [{\citenamefont {De~Pietri}\ \emph {et~al.}(2020)\citenamefont
  {De~Pietri}, \citenamefont {Feo}, \citenamefont {Font}, \citenamefont
  {L\"offler}, \citenamefont {Pasquali},\ and\ \citenamefont
  {Stergioulas}}]{DePietri:2019mti}%
  \BibitemOpen
  \bibfield  {author} {\bibinfo {author} {\bibfnamefont {R.}~\bibnamefont
  {De~Pietri}}, \bibinfo {author} {\bibfnamefont {A.}~\bibnamefont {Feo}},
  \bibinfo {author} {\bibfnamefont {J.~A.}\ \bibnamefont {Font}}, \bibinfo
  {author} {\bibfnamefont {F.}~\bibnamefont {L\"offler}}, \bibinfo {author}
  {\bibfnamefont {M.}~\bibnamefont {Pasquali}}, \ and\ \bibinfo {author}
  {\bibfnamefont {N.}~\bibnamefont {Stergioulas}},\ }\href {\doibase
  10.1103/PhysRevD.101.064052} {\bibfield  {journal} {\bibinfo  {journal}
  {Phys. Rev. D}\ }\textbf {\bibinfo {volume} {101}},\ \bibinfo {pages}
  {064052} (\bibinfo {year} {2020})},\ \Eprint
  {http://arxiv.org/abs/1910.04036} {arXiv:1910.04036 [gr-qc]} \BibitemShut
  {NoStop}%
\bibitem [{\citenamefont {East}\ \emph {et~al.}(2016)\citenamefont {East},
  \citenamefont {Paschalidis},\ and\ \citenamefont {Pretorius}}]{East2016}%
  \BibitemOpen
  \bibfield  {author} {\bibinfo {author} {\bibfnamefont {W.~E.}\ \bibnamefont
  {East}}, \bibinfo {author} {\bibfnamefont {V.}~\bibnamefont {Paschalidis}}, \
  and\ \bibinfo {author} {\bibfnamefont {F.}~\bibnamefont {Pretorius}},\ }\href
  {\doibase 10.1088/0264-9381/33/24/244004} {\bibfield  {journal} {\bibinfo
  {journal} {Classical and Quantum Gravity}\ }\textbf {\bibinfo {volume}
  {33}},\ \bibinfo {pages} {244004} (\bibinfo {year} {2016})}\BibitemShut
  {NoStop}%
\bibitem [{\citenamefont {Radice}\ \emph {et~al.}(2016)\citenamefont {Radice},
  \citenamefont {Bernuzzi},\ and\ \citenamefont {Ott}}]{Radice2016}%
  \BibitemOpen
  \bibfield  {author} {\bibinfo {author} {\bibfnamefont {D.}~\bibnamefont
  {Radice}}, \bibinfo {author} {\bibfnamefont {S.}~\bibnamefont {Bernuzzi}}, \
  and\ \bibinfo {author} {\bibfnamefont {C.~D.}\ \bibnamefont {Ott}},\ }\href
  {\doibase 10.1103/physrevd.94.064011} {\bibfield  {journal} {\bibinfo
  {journal} {Physical Review D}\ }\textbf {\bibinfo {volume} {94}} (\bibinfo
  {year} {2016}),\ 10.1103/physrevd.94.064011}\BibitemShut {NoStop}%
\bibitem [{\citenamefont {Lehner}\ \emph {et~al.}(2016)\citenamefont {Lehner},
  \citenamefont {Liebling}, \citenamefont {Palenzuela},\ and\ \citenamefont
  {Motl}}]{Lehner2016}%
  \BibitemOpen
  \bibfield  {author} {\bibinfo {author} {\bibfnamefont {L.}~\bibnamefont
  {Lehner}}, \bibinfo {author} {\bibfnamefont {S.~L.}\ \bibnamefont
  {Liebling}}, \bibinfo {author} {\bibfnamefont {C.}~\bibnamefont
  {Palenzuela}}, \ and\ \bibinfo {author} {\bibfnamefont {P.~M.}\ \bibnamefont
  {Motl}},\ }\href {\doibase 10.1103/physrevd.94.043003} {\bibfield  {journal}
  {\bibinfo  {journal} {Physical Review D}\ }\textbf {\bibinfo {volume} {94}}
  (\bibinfo {year} {2016}),\ 10.1103/physrevd.94.043003}\BibitemShut {NoStop}%
\bibitem [{\citenamefont {Bustillo}\ \emph
  {et~al.}(2021{\natexlab{a}})\citenamefont {Bustillo}, \citenamefont {Leong},
  \citenamefont {Dietrich},\ and\ \citenamefont {Lasky}}]{CaldernBustillo2021}%
  \BibitemOpen
  \bibfield  {author} {\bibinfo {author} {\bibfnamefont {J.~C.}\ \bibnamefont
  {Bustillo}}, \bibinfo {author} {\bibfnamefont {S.~H.~W.}\ \bibnamefont
  {Leong}}, \bibinfo {author} {\bibfnamefont {T.}~\bibnamefont {Dietrich}}, \
  and\ \bibinfo {author} {\bibfnamefont {P.~D.}\ \bibnamefont {Lasky}},\ }\href
  {\doibase 10.3847/2041-8213/abf502} {\bibfield  {journal} {\bibinfo
  {journal} {The Astrophysical Journal Letters}\ }\textbf {\bibinfo {volume}
  {912}},\ \bibinfo {pages} {L10} (\bibinfo {year}
  {2021}{\natexlab{a}})}\BibitemShut {NoStop}%
\bibitem [{\citenamefont {Lange}\ \emph {et~al.}(2017)\citenamefont {Lange}
  \emph {et~al.}}]{Lange:2017wki}%
  \BibitemOpen
  \bibfield  {author} {\bibinfo {author} {\bibfnamefont {J.}~\bibnamefont
  {Lange}} \emph {et~al.},\ }\href {\doibase 10.1103/PhysRevD.96.104041}
  {\bibfield  {journal} {\bibinfo  {journal} {Phys. Rev.}\ }\textbf {\bibinfo
  {volume} {D96}},\ \bibinfo {pages} {104041} (\bibinfo {year} {2017})},\
  \Eprint {http://arxiv.org/abs/1705.09833} {arXiv:1705.09833 [gr-qc]}
  \BibitemShut {NoStop}%
\bibitem [{\citenamefont {{Shibata}}\ and\ \citenamefont
  {{Kiuchi}}(2017)}]{Shibata-viscous}%
  \BibitemOpen
  \bibfield  {author} {\bibinfo {author} {\bibfnamefont {M.}~\bibnamefont
  {{Shibata}}}\ and\ \bibinfo {author} {\bibfnamefont {K.}~\bibnamefont
  {{Kiuchi}}},\ }\href {\doibase 10.1103/PhysRevD.95.123003} {\bibfield
  {journal} {\bibinfo  {journal} {\prd}\ }\textbf {\bibinfo {volume} {95}},\
  \bibinfo {eid} {123003} (\bibinfo {year} {2017})},\ \Eprint
  {http://arxiv.org/abs/1705.06142} {arXiv:1705.06142 [astro-ph.HE]}
  \BibitemShut {NoStop}%
\bibitem [{\citenamefont {Ruiz}\ \emph
  {et~al.}(2021{\natexlab{b}})\citenamefont {Ruiz}, \citenamefont {Tsokaros},\
  and\ \citenamefont {Shapiro}}]{Ruiz:2021qmm}%
  \BibitemOpen
  \bibfield  {author} {\bibinfo {author} {\bibfnamefont {M.}~\bibnamefont
  {Ruiz}}, \bibinfo {author} {\bibfnamefont {A.}~\bibnamefont {Tsokaros}}, \
  and\ \bibinfo {author} {\bibfnamefont {S.~L.}\ \bibnamefont {Shapiro}},\
  }\href {\doibase 10.1103/PhysRevD.104.124049} {\bibfield  {journal} {\bibinfo
   {journal} {Phys. Rev. D}\ }\textbf {\bibinfo {volume} {104}},\ \bibinfo
  {pages} {124049} (\bibinfo {year} {2021}{\natexlab{b}})},\ \Eprint
  {http://arxiv.org/abs/2110.11968} {arXiv:2110.11968 [astro-ph.HE]}
  \BibitemShut {NoStop}%
\bibitem [{\citenamefont {Chabanov}\ and\ \citenamefont
  {Rezzolla}(2023)}]{Chabanov:2023blf}%
  \BibitemOpen
  \bibfield  {author} {\bibinfo {author} {\bibfnamefont {M.}~\bibnamefont
  {Chabanov}}\ and\ \bibinfo {author} {\bibfnamefont {L.}~\bibnamefont
  {Rezzolla}},\ }\href@noop {} {\  (\bibinfo {year} {2023})},\ \Eprint
  {http://arxiv.org/abs/2307.10464} {arXiv:2307.10464 [gr-qc]} \BibitemShut
  {NoStop}%
\bibitem [{\citenamefont {Bustillo}\ \emph
  {et~al.}(2021{\natexlab{b}})\citenamefont {Bustillo}, \citenamefont {Lasky},\
  and\ \citenamefont {Thrane}}]{Bustillo2021}%
  \BibitemOpen
  \bibfield  {author} {\bibinfo {author} {\bibfnamefont {J.~C.}\ \bibnamefont
  {Bustillo}}, \bibinfo {author} {\bibfnamefont {P.~D.}\ \bibnamefont {Lasky}},
  \ and\ \bibinfo {author} {\bibfnamefont {E.}~\bibnamefont {Thrane}},\ }\href
  {\doibase 10.1103/physrevd.103.024041} {\bibfield  {journal} {\bibinfo
  {journal} {Physical Review D}\ }\textbf {\bibinfo {volume} {103}} (\bibinfo
  {year} {2021}{\natexlab{b}}),\ 10.1103/physrevd.103.024041}\BibitemShut
  {NoStop}%
\bibitem [{\citenamefont {Etienne}\ \emph {et~al.}(2012)\citenamefont
  {Etienne}, \citenamefont {Liu}, \citenamefont {Paschalidis},\ and\
  \citenamefont {Shapiro}}]{Etienne:2011ea}%
  \BibitemOpen
  \bibfield  {author} {\bibinfo {author} {\bibfnamefont {Z.~B.}\ \bibnamefont
  {Etienne}}, \bibinfo {author} {\bibfnamefont {Y.~T.}\ \bibnamefont {Liu}},
  \bibinfo {author} {\bibfnamefont {V.}~\bibnamefont {Paschalidis}}, \ and\
  \bibinfo {author} {\bibfnamefont {S.~L.}\ \bibnamefont {Shapiro}},\ }\href
  {\doibase 10.1103/PhysRevD.85.064029} {\bibfield  {journal} {\bibinfo
  {journal} {Phys. Rev. D}\ }\textbf {\bibinfo {volume} {85}},\ \bibinfo
  {pages} {064029} (\bibinfo {year} {2012})},\ \Eprint
  {http://arxiv.org/abs/1112.0568} {arXiv:1112.0568 [astro-ph.HE]} \BibitemShut
  {NoStop}%
\bibitem [{\citenamefont {Ruiz}\ \emph
  {et~al.}(2020{\natexlab{a}})\citenamefont {Ruiz}, \citenamefont
  {Paschalidis}, \citenamefont {Tsokaros},\ and\ \citenamefont
  {Shapiro}}]{Ruiz:2020elr}%
  \BibitemOpen
  \bibfield  {author} {\bibinfo {author} {\bibfnamefont {M.}~\bibnamefont
  {Ruiz}}, \bibinfo {author} {\bibfnamefont {V.}~\bibnamefont {Paschalidis}},
  \bibinfo {author} {\bibfnamefont {A.}~\bibnamefont {Tsokaros}}, \ and\
  \bibinfo {author} {\bibfnamefont {S.~L.}\ \bibnamefont {Shapiro}},\ }\href
  {\doibase 10.1103/PhysRevD.102.124077} {\bibfield  {journal} {\bibinfo
  {journal} {Phys. Rev. D}\ }\textbf {\bibinfo {volume} {102}},\ \bibinfo
  {pages} {124077} (\bibinfo {year} {2020}{\natexlab{a}})},\ \Eprint
  {http://arxiv.org/abs/2011.08863} {arXiv:2011.08863 [astro-ph.HE]}
  \BibitemShut {NoStop}%
\bibitem [{\citenamefont {Ruiz}\ \emph
  {et~al.}(2020{\natexlab{b}})\citenamefont {Ruiz}, \citenamefont {Tsokaros},\
  and\ \citenamefont {Shapiro}}]{Ruiz:2020via}%
  \BibitemOpen
  \bibfield  {author} {\bibinfo {author} {\bibfnamefont {M.}~\bibnamefont
  {Ruiz}}, \bibinfo {author} {\bibfnamefont {A.}~\bibnamefont {Tsokaros}}, \
  and\ \bibinfo {author} {\bibfnamefont {S.~L.}\ \bibnamefont {Shapiro}},\
  }\href {\doibase 10.1103/PhysRevD.101.064042} {\bibfield  {journal} {\bibinfo
   {journal} {Phys. Rev. D}\ }\textbf {\bibinfo {volume} {101}},\ \bibinfo
  {pages} {064042} (\bibinfo {year} {2020}{\natexlab{b}})},\ \Eprint
  {http://arxiv.org/abs/2001.09153} {arXiv:2001.09153 [astro-ph.HE]}
  \BibitemShut {NoStop}%
\bibitem [{\citenamefont {Baiotti}\ \emph {et~al.}(2008)\citenamefont
  {Baiotti}, \citenamefont {Giacomazzo},\ and\ \citenamefont
  {Rezzolla}}]{Baiotti:2008ra}%
  \BibitemOpen
  \bibfield  {author} {\bibinfo {author} {\bibfnamefont {L.}~\bibnamefont
  {Baiotti}}, \bibinfo {author} {\bibfnamefont {B.}~\bibnamefont {Giacomazzo}},
  \ and\ \bibinfo {author} {\bibfnamefont {L.}~\bibnamefont {Rezzolla}},\
  }\href {\doibase 10.1103/PhysRevD.78.084033} {\bibfield  {journal} {\bibinfo
  {journal} {Phys. Rev. D}\ }\textbf {\bibinfo {volume} {78}},\ \bibinfo
  {pages} {084033} (\bibinfo {year} {2008})},\ \Eprint
  {http://arxiv.org/abs/0804.0594} {arXiv:0804.0594 [gr-qc]} \BibitemShut
  {NoStop}%
\bibitem [{\citenamefont {Raithel}\ and\ \citenamefont
  {Paschalidis}(2023)}]{Raithel:2023zml}%
  \BibitemOpen
  \bibfield  {author} {\bibinfo {author} {\bibfnamefont {C.~A.}\ \bibnamefont
  {Raithel}}\ and\ \bibinfo {author} {\bibfnamefont {V.}~\bibnamefont
  {Paschalidis}},\ }\href {\doibase 10.1103/PhysRevD.108.083029} {\bibfield
  {journal} {\bibinfo  {journal} {Phys. Rev. D}\ }\textbf {\bibinfo {volume}
  {108}},\ \bibinfo {pages} {083029} (\bibinfo {year} {2023})},\ \Eprint
  {http://arxiv.org/abs/2306.13144} {arXiv:2306.13144 [astro-ph.HE]}
  \BibitemShut {NoStop}%
\bibitem [{\citenamefont {Rivieccio}\ \emph {et~al.}(2024)\citenamefont
  {Rivieccio}, \citenamefont {Guerra}, \citenamefont {Ruiz},\ and\
  \citenamefont {Font}}]{Rivieccio:2024sfm}%
  \BibitemOpen
  \bibfield  {author} {\bibinfo {author} {\bibfnamefont {G.}~\bibnamefont
  {Rivieccio}}, \bibinfo {author} {\bibfnamefont {D.}~\bibnamefont {Guerra}},
  \bibinfo {author} {\bibfnamefont {M.}~\bibnamefont {Ruiz}}, \ and\ \bibinfo
  {author} {\bibfnamefont {J.~A.}\ \bibnamefont {Font}},\ }\href {\doibase
  10.1103/PhysRevD.109.064032} {\bibfield  {journal} {\bibinfo  {journal}
  {Phys. Rev. D}\ }\textbf {\bibinfo {volume} {109}},\ \bibinfo {pages}
  {064032} (\bibinfo {year} {2024})},\ \Eprint
  {http://arxiv.org/abs/2401.06849} {arXiv:2401.06849 [astro-ph.HE]}
  \BibitemShut {NoStop}%
\bibitem [{\citenamefont {Takami}\ \emph {et~al.}(2015)\citenamefont {Takami},
  \citenamefont {Rezzolla},\ and\ \citenamefont {Baiotti}}]{Takami:2014tva}%
  \BibitemOpen
  \bibfield  {author} {\bibinfo {author} {\bibfnamefont {K.}~\bibnamefont
  {Takami}}, \bibinfo {author} {\bibfnamefont {L.}~\bibnamefont {Rezzolla}}, \
  and\ \bibinfo {author} {\bibfnamefont {L.}~\bibnamefont {Baiotti}},\ }\href
  {\doibase 10.1103/PhysRevD.91.064001} {\bibfield  {journal} {\bibinfo
  {journal} {Phys. Rev. D}\ }\textbf {\bibinfo {volume} {91}},\ \bibinfo
  {pages} {064001} (\bibinfo {year} {2015})},\ \Eprint
  {http://arxiv.org/abs/1412.3240} {arXiv:1412.3240 [gr-qc]} \BibitemShut
  {NoStop}%
\bibitem [{\citenamefont {Bamber}\ \emph {et~al.}(2025)\citenamefont {Bamber},
  \citenamefont {Tsokaros}, \citenamefont {Ruiz},\ and\ \citenamefont
  {Shapiro}}]{Bamber:2024qzi}%
  \BibitemOpen
  \bibfield  {author} {\bibinfo {author} {\bibfnamefont {J.}~\bibnamefont
  {Bamber}}, \bibinfo {author} {\bibfnamefont {A.}~\bibnamefont {Tsokaros}},
  \bibinfo {author} {\bibfnamefont {M.}~\bibnamefont {Ruiz}}, \ and\ \bibinfo
  {author} {\bibfnamefont {S.~L.}\ \bibnamefont {Shapiro}},\ }\href {\doibase
  10.1103/PhysRevD.111.044038} {\bibfield  {journal} {\bibinfo  {journal}
  {Phys. Rev. D}\ }\textbf {\bibinfo {volume} {111}},\ \bibinfo {pages}
  {044038} (\bibinfo {year} {2025})},\ \Eprint
  {http://arxiv.org/abs/2411.00943} {arXiv:2411.00943 [gr-qc]} \BibitemShut
  {NoStop}%
\end{thebibliography}%
%
\end{document}